\documentclass[floatfix,amsmath,amssymb,showpacs,showkeys,preprintnumbers,nofootinbib,sort&compress,prd]{revtex4}%
\usepackage{float,amstext,amsthm,slashed,epigraph}
\usepackage{amsfonts}
\usepackage{amsmath}
\usepackage{graphicx}
\usepackage{dcolumn}
\usepackage{bm}
\usepackage{textcomp}
\usepackage{amssymb}%
\usepackage{bbm}
\usepackage[usenames]{color}

\begin{document}
\title{Hadronic decays of the (pseudo-)scalar charmonium states $\eta_c$ and $\chi_{c0}$ in the 
extended Linear Sigma Model}
\author{Walaa I.\ Eshraim$^{1,2}$ and Christian S. Fischer$^{1}$ }
\affiliation{$^{1}$Institut f\"ur Theoretische Physik, Justus-Liebig-Universit\"at Giessen, 35392 Giessen, Germany}
\affiliation{$^{2}$Helmholtz Institute Mainz, Staudingerweg 18,
55128 Mainz, Germany}

\begin{abstract}
We study the phenomenology of the ground-state (pseudo-)scalar charmonia $\eta_c$ and $\chi_{c0}$ in the 
framework of a $U(4)_r \times U(4)_l$ symmetric linear sigma model with (pseudo-)scalar and (axial-) vector 
mesons. Based on previous results for the spectrum of charmonia and the spectrum and (OZI-dominant) strong 
decays of open charmed mesons, we extend the study of this model to OZI-suppressed charmonia decays. 
This includes decays into 'ordinary' mesons but also particularly interesting channels with scalar-isoscalar
resonances $f_0(1370),\, f_0(1500),\, f_0(1710)$ that may include sizeable contributions from a scalar glueball. 
We study the variation of the corresponding decay widths assuming different mixings between glueball and
quark-antiquark states. We also compute the decay width of the pseudoscalar $\eta_c$ into a pseudoscalar 
glueball. In general, our results for decay widths are in reasonable agreement with experimental 
data where available. Order of magnitude predictions for as yet unmeasured states and channels are potentially
interesting for BESIII, Belle II, LHCb as well as the future PANDA experiment at the FAIR facility.  
\end{abstract}

\maketitle

\section{Introduction}

In recent years, the charm quark energy region has been the focus of many theoretical and experimental 
investigations \cite{Patrignani:2016xqp}. In particular the plethora of recently discovered exotic states,
i.e. states with quark content beyond $q\bar{q}$ and $qqq$ or states with quantum numbers not accounted 
for in non-relativistic quark models, have raised enormous interest. With the copious production
of 'ordinary' charmonia in experiments such as BESIII, Belle (II) and LHCb, however, also their potentially
rare decays into light hadrons gains interest. These transitions occur via gluon-rich processes and are 
therefore interesting with respect to the interplay and transition of perturbative and non-perturbative QCD.
Consequently, theoretical work in the past has concentrated on the application and limits of perturbative QCD
and the construction of non-perturbative models for the (OZI-suppressed) production of $q\bar{q}$-pairs
in the decays, see e.g. \cite{Bodwin:1992ye,Zhou:2004mw,Close:2005vf,Zhao:2007ze,Segovia:2013kg,Frere:2015xxa} 
and Refs. therein.   

Exotic states, however not only occur in the charm quark region, but may be present already in the light 
quark sector. In particular the light scalar meson sector is discussed frequently, with a putative scalar 
isoscalar glueball in the energy region around 1.5 GeV. The decay of charmonia into pairs of these states
may offer the interesting possibility to study the nature of these states using information from the 
gluon-rich decay processes. 

In order to study these decays we employ the extended Linear Sigma Model (eLSM) \cite{Ko:1994en,Urban:2001ru},
an effective model describing the vacuum phenomenology of (pseudo-)scalar and (axial-)vector mesons in the 
cases of $N_f=2$ \cite{Parganlija:2010fz, Gallas:2009qp, Janowski:2011gt} and $N_f=3$ \cite{Parganlija:2012fy}. 
Recently, the framework has been generalised to $N_f=4$ 
\cite{Eshraim:2014afa,Eshraim:2014vfa,Eshraim:2014gya,Eshraim:2014tla,Eshraim:2014eka,Eshraim:2015cia} 
and applied to the calculation of masses of open and hidden charmed mesons as well as the decays of open 
charmed mesons in the low energy limit. Although the model is rooted in chiral symmetry and its breaking 
pattern in the light quark sector, these applications have been (perhaps unexpectedly) quite successful. 
In this work we extend these studies to also include the decays of hidden charm mesons. 

The construction of the eLSM is based on a global chiral symmetry $U(N_f)_r \times U(N_f)_l$ as well as the 
classical dilation symmetry. In the vacuum, global chiral symmetry is broken spontaneously by a non-vanishing
expectation value of the quark condensate, anomalously by quantum effects, and explicitly by non-vanishing 
quark masses. Furthermore, the dilation symmetry is broken explicitly. As a consequence, besides the 
usual meson multiplets the eLSM also includes two glueballs composed of two gluons each: 
a scalar one (denoted as $G$) \cite{Rosenzweig:1981cu,Migdal:1982jp,Gomm:1985ut,Gomm:1984zq} and 
a pseudoscalar one (denoted as $\widetilde{G}$). 
The identification of the scalar glueballs with experimentally observed states is notoriously complicated, see e.g. 
\cite{Amsler:1995td,Lee:1999kv,Close:2001ga,Amsler:2004ps,Giacosa:2005zt,Giacosa:2005bw,Cheng:2006hu,Klempt:2007cp,Chatzis:2011qz} 
for discussions. In general, there is a mixing between states: the non-strange
$\sigma_N\equiv(u\overline{u}+d\overline{d})/\sqrt{2}$, hidden-strange $\sigma_S\equiv s\overline{s}$, and 
scalar glueball $G\equiv gg$ \cite{Close:2005vf,Zhao:2007ze,Frere:2015xxa}. 
Within the eLSM this three-body mixing in the scalar-isoscalar channel was 
resolved in Ref.~\cite{Janowski:2011gt,Janowski:2014ppa} and generated the physical resonances 
$f_0(1370),\,f_0(1500)$ and $f_0(1710)$. Of these, the $f_0(1710)$ has the largest overlap with the 
scalar glueball. Its counterpart in the pseudoscalar sector has been studied within the eLSM in 
Refs. \cite{Eshraim:2012jv,Eshraim:2012rb,Eshraim:2012ju,Eshraim:2013dn,Eshraim:2016mds}. 
In the charm sector, the eLSM contains so far four charmonium states, which are (pseudo-)scalar and 
(axial-)vector ground states $\eta_c,\chi_{c0},J/\psi(1S)$ and $\chi_{c1}$ \cite{Eshraim:2014eka}.
In the present work we study the OZI-suppressed decays of the two (pseudo-)scalar charmonia\footnote{Note 
that the current set-up of the eLMS does not contain decay channels of the two (axial-)vector charmonia.
These, however, have been considered in other approaches, see e.g. \cite{Geng:2009iw,Dai:2015cwa} 
and Refs. therein.}. 

We use eLSM parameters in the light, strange and charm quark sectors fixed previously
\cite{Parganlija:2012fy,Eshraim:2014eka}. In addition we introduce two new parameters, $\lambda_1^C,\,h_1^C$, 
which control the decay widths of $\eta_c$ and $\chi_{c0}$. This, together
with a summary of the construction of the eLSM is discussed in section \ref{secII}. In section \ref{secIII} we
present the results of the two- and three-body decay widths of $\eta_c$ and $\chi_{c0}$, and discuss their
significance. In section \ref{secIV} we conclude. Details of the calculations are relegated to the Appendices. 

\section{The $U(4)_r \times U(4)_l$ LSM interaction with glueballs}\label{secII}

The $U(4)_{L}\times U(4)_{R}$ linear sigma model with (pseudo)scalar and (axial-)vector mesons, 
a scalar $G$ and a pseudoscalar glueball $\widetilde{G}$ is given by \cite{Eshraim:2014eka}

\begin{align}
\mathcal{L}  &
=\mathcal{L}_{dil}+\mathrm{Tr}[(D^{\mu}\Phi)^{\dagger}(D^{\mu
}\Phi)]-m_{0}^{2}\left(  \frac{G}{G_{0}}\right)  ^{2}\mathrm{Tr}(\Phi^{\dagger}%
\Phi)
%-\lambda_{1}[\mathrm{Tr}(\Phi^{\dagger}\Phi)]^{2}
-\lambda_{1}^S[\mathrm{Tr}((\mathbbm{1}-\mathbb{P}_C)\Phi^{\dagger}(\mathbbm{1}-\mathbb{P}_C)\Phi)]^{2}
-\lambda_{1}^C[\mathrm{Tr}(\Phi^{\dagger}\Phi)]^{2}
\nonumber\\
& -\lambda_{2}\mathrm{Tr}(\Phi^{\dagger}\Phi)^{2} +\mathrm{Tr}\left\{  \left[  \left(  \frac{G}{G_{0}}\right)
^{2}\frac {m_{1}^{2}}{2}+\Delta\right]  \left[
(L^{\mu})^{2}+(R^{\mu})^{2}\right] \right\}
-\frac{1}{4}\mathrm{Tr}[(L^{\mu\nu})^{2}+(R^{\mu\nu})^{2}]
-2\,\mathrm{Tr}[\varepsilon
\Phi^{\dagger}\Phi]\nonumber\\
&+\mathrm{Tr}[H(\Phi+\Phi^{\dagger})]+c(det\Phi-det\Phi^{\dagger})^{2}-
\delta \widetilde{c}(det\Phi-det\Phi^{\dagger})^{2}\mathrm{Tr}(\mathbb{P}_C\Phi^{\dagger}\mathbb{P}_C\Phi)+i\tilde{c}\,\tilde{G}\left(
\text{\textrm{det}}\Phi-\text{\textrm{det}}\Phi^{\dag}\right)\nonumber\\
&
%+\frac{h_{1}}{2}\mathrm{Tr}(\Phi^{\dagger}\Phi)Tr[(L^{\mu})^{2}+(R^{\mu})^{2}]
+\frac{h_{1}^S}{2}\mathrm{Tr}((\mathbbm{1}-\mathbb{P}_C)\Phi^{\dagger}(\mathbbm{1}-\mathbb{P}_C)\Phi)\mathrm{Tr}[(L^{\mu})^{2}+(R^{\mu})^{2}]
+\frac{h_{1}^C}{2}\mathrm{Tr}(\Phi^{\dagger}\Phi)\mathrm{Tr}[(L^{\mu})^{2}+(R^{\mu})^{2}] \nonumber\\
&+h_{2}\mathrm{Tr}[(\Phi R^{\mu})^{2}+(L^{\mu}\Phi)^{2}] +2h_{3}%
\mathrm{Tr}(\Phi R_{\mu}\Phi^{\dagger}L^{\mu})\nonumber\\
&
+i\frac{g_{2}}{2}\{\mathrm{Tr}(L_{\mu\nu}[L^{\mu},L^{\nu}])+\mathrm{Tr}(R_{\mu\nu
}[R^{\mu},R^{\nu}])\}+...\text{ ,} \label{Lagc}%
\end{align}
where 
$D^\mu\Phi\equiv\partial^\mu\Phi-ig_1 (L^\mu \Phi-\Phi R^\mu)$ is the covariant derivative; 
$L^{\mu\nu}\equiv \partial^\mu L^\nu -\partial^\nu L^\mu$, and $R^{\mu\nu}\equiv\partial^\mu R^\nu -\partial^\nu R^\mu$ 
are the left-handed and right-handed field strength tensors. In addition to chiral symmetry it features
dilatation invariance and invariance under the discrete symmetries $C$ and $P$. In the following we 
summarise the most important features of the so defined extended linear sigma model (eLSM). 

In our framework with $N_f=4$ quark flavours the field $\Phi$ represents the $4\times4$ (pseudo)scalar multiplets 
\begin{equation}\label{4.1}
\Phi=(S^{a}+iP^{a})t^{a}=\frac{1}{\sqrt{2}}
\left(%
\begin{array}{cccc}
  \frac{(\sigma_{N}+a^0_{0})+i(\eta_N +\pi^0)}{\sqrt{2}} & a^{+}_{0}+i \pi^{+} & K^{*+}_{0}+iK^{+} & D^{*0}_0+iD^0 \\
  a^{-}_{0}+i \pi^{-} & \frac{(\sigma_{N}-a^0_{0})+i(\eta_N -\pi^0)}{\sqrt{2}} & K^{*0}_{0}+iK^{0} & D^{*-}_0+iD^{-} \\
  K^{*-}_{0}+iK^{-} & \overline{K}^{*0}_{0}+i\overline{K}^{0} & \sigma_{S}+i\eta_{S} & D^{*-}_{S0}+iD^{-}_S\\
  \overline{D}^{*0}_0+i\overline{D}^0 & D^{*+}_0+iD^{+} & D^{*+}_{S0}+iD^{+}_S & \chi_{C0}+i\eta_C\\
\end{array}%
\right),
\end{equation}
where $t^{a}$ are the generators of the group $U(N_{f})$. The multiplet $\Phi$ transforms 
as $\Phi\rightarrow U_{L}\Phi U_{R}^{\dagger}$ under $U_{L}(4)\times U_{R}(4)$ chiral
transformations with $U_{L(R)}=e^{-i\theta_{L(R)}^at^a}$, 
with $\Phi(t,\vec{x})\rightarrow\Phi^{\dagger}(t,-\vec{x})$ under parity transformations,
and $\Phi\rightarrow\Phi^{\dagger}$ under charge conjugation. 
The determinant of $\Phi$ is invariant under $SU(4)_{L} \times SU(4)_{R}$, but not under $U(1)_{A}$ 
because ${\rm det \Phi}\rightarrow {\rm det} U_{A}\Phi U_A=e^{-i\theta_{A}^0\sqrt{2N_f}}{\rm det \Phi}\neq {\rm det \Phi}$.

Next we present the left- and right-handed (axial)vector multiplets $L^\mu$ and $R^\mu$ 
containing the vector and axial-vector degrees of freedom $V^{a}$ and $A^{a}$ 
\begin{equation}\label{4.2}
L^\mu=(V^a+i\,A^a)^{\mu}\,t^a=\frac{1}{\sqrt{2}}
\left(%
\begin{array}{cccc}
  \frac{\omega_N+\rho^{0}}{\sqrt{2}}+ \frac{f_{1N}+a_1^{0}}{\sqrt{2}} & \rho^{+}+a^{+}_1 & K^{*+}+K^{+}_1 & D^{*0}+D^{0}_1 \\
  \rho^{-}+ a^{-}_1 &  \frac{\omega_N-\rho^{0}}{\sqrt{2}}+ \frac{f_{1N}-a_1^{0}}{\sqrt{2}} & K^{*0}+K^{0}_1 & D^{*-}+D^{-}_1 \\
  K^{*-}+K^{-}_1 & \overline{K}^{*0}+\overline{K}^{0}_1 & \omega_{S}+f_{1S} & D^{*-}_{S}+D^{-}_{S1}\\
  \overline{D}^{*0}+\overline{D}^{0}_1 & D^{*+}+D^{+}_1 & D^{*+}_{S}+D^{+}_{S1} & J/\psi+\chi_{C1}\\
\end{array}%
\right)^\mu\,,
\end{equation}
$$\\$$
\begin{equation}\label{5}
R^\mu=(V^a-i\,A^a)^\mu\,t^a=\frac{1}{\sqrt{2}}
\left(%
\begin{array}{cccc}
  \frac{\omega_N+\rho^{0}}{\sqrt{2}}- \frac{f_{1N}+a_1^{0}}{\sqrt{2}} & \rho^{+}-a^{+}_1 & K^{*+}-K^{+}_1 & D^{*0}-D^{0}_1 \\
  \rho^{-}- a^{-}_1 &  \frac{\omega_N-\rho^{0}}{\sqrt{2}}-\frac{f_{1N}-a_1^{0}}{\sqrt{2}} & K^{*0}-K^{0}_1 & D^{*-}-D^{-}_1 \\
  K^{*-}-K^{-}_1 & \overline{K}^{*0}-\overline{K}^{0}_1 & \omega_{S}-f_{1S} & D^{*-}_{S}-D^{-}_{S1}\\
  \overline{D}^{*0}-\overline{D}^{0}_1 & D^{*+}-D^{+}_1 & D^{*+}_{S}-D^{+}_{S1} & J/\psi-\chi_{C1}\\
\end{array}%
\right)^\mu\,,
\end{equation}
which transform as $L^{\mu}\rightarrow U_{L} L^{\mu}U_{L}^{\dag}$ and $R^{\mu}\rightarrow U_{R}L^{\mu}U_{R}^{\dag}$ 
under chiral transformations. These transformation properties of $\Phi,\, L^\mu$, and $R^\mu$ have
been used to build the chirally invariant Lagrangian (\ref{Lagc}).

If $m_0^2<0$, the Lagrangian (\ref{Lagc}) also includes the effects of spontaneous symmetry breaking. 
To make this apparent one shifts the scalar-isoscalar fields
$G,\sigma_{N},\,\sigma_{S},$ and $\chi_{C0}$ by their vacuum
expectation values $G_{0},\,\phi_{N},\,\phi_{S},$ and $\phi_{C}$
\cite{Janowski:2011gt, Eshraim:2014eka}
\begin{align}
G\rightarrow
G+G_{0},\,\,\,\sigma_N\rightarrow\sigma_N+\phi_N\,,\,\,\,\,\,\nonumber\\
\sigma_S\rightarrow\sigma_S+\phi_S\;,\,\,\,
\chi_{C0}\rightarrow\chi_{C0}+\phi_C\;,
\end{align}
where $\phi_N,\,\phi_S$ and $\phi_C$ are the corresponding chiral shifts, which read
$$\phi_N=164.6 \,\,\rm{MeV},\,\,\,\phi_S=126.2 \,\,\rm{MeV} \,\,\, \phi_C=176 \,\,\rm{MeV}\,. $$
In order to be consistent with the full effective chiral Lagrangian of the extended 
Linear Sigma Model \cite{Eshraim:2014eka} we also shift the axial-vector fields and 
thus redefine the wave functions of the pseudoscalar fields
\begin{align}
&\pi^{\pm,0}\rightarrow
Z_{\pi}\pi^{\pm,0},\,\,\,\,\,\,\,\,\,\,\,\,\,\,\,\,\,\,\,\,\,\,\,\,\,\,\,\,\,\,\,\,\,\,\,\,\,\,\,\,\,\,\,\,K^{\pm,0,\bar{0}}\rightarrow
Z_{K}K^{\pm,0,\bar{0}}\text{ },\nonumber\\
&\eta_{N/S/C}\rightarrow Z_{\eta_{N}
/\eta_{S}/\eta_{C}}\eta_{N/S/C},\,\,\,\,\,\,\,\,\,\,{K^{\star}_0}^{\pm,0,\bar{0}}\rightarrow
Z_{K^{\star}}{K^{\star}_0}^{\pm,0,\bar{0}}\,,\label{rensns}\\
&D^{\pm,0,\bar{0}}\rightarrow
Z_{D}D^{\pm,0,\bar{0}},\,\,\,\,\,\,\,\,\,\,\,\,\,\,\,\,\,\,\,\,\,\,\,\,\,\,\,\,\,\,\,\,D^{*\pm}_{0}\rightarrow
Z_{D^{*}_{0}}D^{*\pm}_{0}\>,\nonumber\\
&D_{0}^{\ast0,\bar{0}}\rightarrow Z_{D_{0}^{\ast0}}D_{0}^{\ast0,\bar{0}}%
,\,\,\,\,\,\,\,\,\,\,\,\,\,\,\,\,\,\,\,\,\,\,\,\,\,\,\,\,\,\,\,\,D_{S0}^{\ast\pm}\rightarrow
Z_{D_{S0}^{\ast}}D_{S0}^{\ast\pm
}\text{ ,}\label{renc}%
\end{align}
where $Z_i$ are the renormalization constants of the corresponding
wave functions \cite{Eshraim:2014eka}. 

The terms involving the matrices $H$, $\epsilon$ and $\Delta$ correspond to explicit breaking of the dilaton
and chiral symmetry due to non-zero current quark masses. They are all diagonal with constant entries (see
\cite{Eshraim:2014eka} for details):
\begin{align}
H           & =\frac{1}{2}diag(h_{N},\, h_{N},\,\sqrt{2}h_{S},\,\sqrt{2}h_{C})\\
\Delta      & =           diag(\delta_{N},\,\delta_{N},\,\delta_{S},\,\delta_{C})\\
\varepsilon &=diag(\varepsilon_{N},\,\varepsilon_{N},\,\varepsilon_{S},\,\varepsilon_{C})
\end{align}
where $h_{i}\sim m_i$ $\delta_{i}\sim m_i^2\,\,{\rm and} \,\,\varepsilon_{i}\sim m_i^2$.

In order to make contact with experiment one needs to assign the various fields of the model to physical states:\\ 
(i) In the pseudoscalar sector $P^a$ the fields $\vec{\pi}$ and $\vec{K}$ 
correspond to the physical pion iso-triplet and the kaon iso-doublet, respectively \cite{Janowski:2014ppa}.
The bare fields $\eta_{N} \equiv\left\vert \bar{u}u+\bar{d}d\right\rangle /\sqrt{2}$ and 
$\eta_{S} \equiv\left\vert \bar{s}s\right\rangle $ are the non-strange and strange mixing 
contributions of the physical states $\eta$ and $\eta^{\prime}$ 
with mixing angle $\varphi\simeq-44.6^{\circ}$ \cite{Parganlija:2012fy,Eshraim:2012jv,Janowski:2014ppa}:
\begin{equation}
\eta=\eta_{N}\cos\varphi+\eta_{S}\sin\varphi,\text{
}\eta^{\prime}=-\eta
_{N}\sin\varphi+\eta_{S}\cos\varphi\,. \label{mixetas}%
\end{equation}
In the pseudoscalar charm sector, we have the well-established $D$ resonances, the open strange-charmed 
$D_s$ states, and the charm-anticharm state $\eta_c(1S)$.\\
(ii) In the vector sector $V^a$ the iso-triplet fields
$\vec{\rho}$, the kaon states $\vec{K}^\ast$, and the
isoscalar states $\omega_N$ and $\omega_S$ correspond to the
$\rho(770)$, $K^\ast(892)$, $\omega$, and $\phi$ mesons,
respectively. Notice that the mixing between strange and
non-strange iso-scalars is small. The charm sectors
$D^{\ast0,\overline{0},\pm}$, $D_S^{\ast\pm}$, and charmonium
state $J/\psi$ correspond to the open-charm sectors
$D^\ast(2007)^0,\, D^\ast(2010)^\pm,\,$ $D^{\ast\pm}_s$ (with
mass$= 2112.3\pm0.5$ MeV), and $J/\psi(1S)$,
respectively.\\
(iii) In the axial-vector sector $A^a$, the iso-triplet field
$\vec{a}_1(1260)$, the kaon states $\vec{K}_1$, the isoscalar fields
$f_{1,N}$ and $f_{1,S}$, the open-charm sector $\vec{D}_1$ and the
strange-charmed doublet $\vec{D}^{\pm}_{S1}$ are assigned to
$a_1(1260),\,K_1(1270),$ or $K_1(1400)$ mesons, $f_1(1285),\,
f_1(1420),$ $D_1(2420)^{0,\pm},$ and $D_{S1}(2536)^{\pm}$,
respectively. The charm-anticharm state $\chi_{C,1}$
represent the ground-state charmonium resonance $\chi_{c,1}(1P)$.
For more detail of strange-non-strange fields assignment see Refs.
\cite{Parganlija:2012fy, Giacosa:2009bj,Giacosa:2010by, Amsler:2004ps, Klempt:2007cp} and for open and hidden charm
fields assignment see Refs. \cite{Eshraim:2014eka}.
(iv) In the scalar sector $S^a$ the iso-triplet $\vec{a}_0$ and the kaon $\vec{K}^\ast_0$ fields 
are assigned to the physical states $a_0(1450)$ and $K^\ast_0(1430)$, respectively,
(the details of this assignment are given in Ref.\ \cite{Parganlija:2012fy}). 
The open charmed sector $D^\ast_0$ is assigned to the resonance
$D^\ast_0(2400)$ whereas the strange-charm sector $D^\ast_{S0}$ to
the $D^\ast_{S0}(2317)$ and the charmonium sector $\chi_{C0}$
corresponds to the ground-state charm-anticharm resonance $\chi_{C0}$.

The new element in this work is the calculation of the strong decays of charmonia. As we will discuss 
in detail below, this will be accomplished by the terms in the Lagrangian (\ref{Lagc}) involving the
new parameters $\lambda_1^C$ and $h_1^C$. In some previous works the corresponding parameters $\lambda_1^S$ 
and $h_1^S$ in the light quark sector have either been set to zero
\cite{Parganlija:2010fz,Eshraim:2014eka,Eshraim:2015cia} or have been 
determined from decays of states with light quarks \cite{Parganlija:2012fy,Janowski:2014ppa}. In contrast
to these, the decays in the heavy quark sector always come with charm quark-antiquark annihilations. In
order to reflect the different physics of these processes we use the projection operator 
$\mathbb{P}_C = {\rm diag}\{0,0,0,1\}$ onto the charmed states to introduce separate parameters 
$\lambda_1^C$ and $h_1^C$ in this sector.
In addition, it is 
possible to improve the description of charmonia decays by taking into account interaction terms that 
break chiral symmetry explicitly \cite{Eshraim:2014eka}. The rationale behind this approach is that 
the explicit breaking of chiral symmetry due to the charm quark mass is large anyhow and therefore 
there is no point in maintaining chiral symmetry in the effective operators describing the decays. 
In this work we explore the effect of a corresponding modification of the anomaly term that affects the
decays into pseudoscalar flavour singlet mesons, i.e. we consider 
\begin{equation}
c(det\Phi -det\Phi^{\dagger})^{2}\rightarrow c(det\Phi -det\Phi^{\dagger})^{2} -\delta \widetilde{c}(det\Phi -det\Phi^{\dagger})^{2}  \mathrm{Tr}( \mathbb{P}_C \Phi^{\dagger} \mathbb{P}_C\Phi)\,, \label{delta}
\end{equation}
where $c$ and $\delta \widetilde{c}$ are dimensionful constants. Again, $\mathbb{P}_C = {\rm diag}\{0,0,0,1\}$ 
is a projection operator onto the charmed states that ensures that the light meson sector remains untouched
by our modification.

In addition to the meson fields, the Lagrangian Eq.~(\ref{Lagc}) also contains scalar and pseudoscalar
glueballs. The dilation Lagrangian $\mathcal{L}_{dil}$ describes a scalar glueball built from two gluons 
$G\equiv|gg\rangle$ with quantum number $J^{PC}=0^{++}$ and mimics the trace anomaly of the 
pure Yang-Mills sector of QCD 
\cite{Rosenzweig:1981cu,Migdal:1982jp,Gomm:1985ut,Gomm:1984zq,Parganlija:2012fy}:
\begin{equation}
\mathcal{L}_{dil}=\frac{1}{2}(\partial_{\mu}G)^{2}-\frac{1}{4}\frac{m_{G}^{2}%
}{\Lambda^{2}}\left(
G^{4}\,\log\frac{G}{\Lambda}-\frac{G^{4}}{4}\right)\,.
\label{dil}%
\end{equation}
The energy scale of low-energy QCD is described by the dimensionful parameter $\Lambda$ which is identical to the 
minimum $G_0$ of the dilaton potential ($G_0=\Lambda$). The scalar glueball mass $m_G$ has been evaluated by 
lattice QCD which gives a mass of about (1.5-1.7) GeV
\cite{Bali:1993fb,Morningstar:1999rf,Bali:2000vr,Loan:2005ff,Chen:2005mg,Gregory:2012hu}. 
The dilatation symmetry or scale invariance, $x^\mu\rightarrow\lambda^{-1}x^{\mu}$, is realized at the
classical level of the Yang-Mills sector of QCD and explicitly broken due to the logarithmic term of the 
dilaton potential. This breaking leads to the divergence of the corresponding current:
$\partial_\mu J^\mu_{dil}=T^\mu_{dil,\,\mu}=-\frac{1}{4}m_G^2\Lambda^2$ \cite{Janowski:2011gt}. 

The identification of the scalar (isoscalar) glueball in the experimental spectrum is highly controversial. 
Here, the quark-antiquark states 
$\sigma_{N}\equiv\left\vert \bar{u}u+\bar{d}d\right\rangle /\sqrt{2}=|n\bar{n}\rangle$,
$\sigma_{S}\equiv\left\vert \bar{s}s\right\rangle $, and the scalar glueball $G$ mix and generate the 
three resonances $f_0(1370),\, f_0(1500),\, {\rm and}\, f_0(1710)$. 
In the flavour singlet basis this mixing is expressed via 
\begin{equation}\label{scalmix1}
\left(%
\begin{array}{c}
 |f_0(1710)\rangle \\
 |f_0(1500)\rangle   \\
 |f_0(1370)\rangle\\
\end{array}%
\right)
=U_n\left(%
\begin{array}{c}
 |G\rangle\\
  |s\overline{s}\rangle\\
  |n\overline{n}\rangle\\
\end{array}%
\right);\hspace{2cm}
U_1=\left(%
\begin{array}{ccccc}
0.93	& 0.26	& -0.27	\\
-0.17	& 0.94	& 0.30	\\
-0.33	& 0.24	& -0.91	\\ 
\end{array}%
\right)\,,
\end{equation}
where $U_1$ represents the best result in the extended linear sigma model \cite{Janowski:2014ppa}
from a systematic fit to the spectrum and decay properties of light mesons. Thus the resonance $f_{0}(1710)$ 
is identified as being predominantly a scalar glueball. This matrix will be used below 
in our consistent evaluation of the decays of the $\chi_{C0}$ into scalar isoscalar states. 
However, in order to contrast the resulting decay widths with the ones of potentially different
assignments we will also play with other mixing matrices. Since the charm quark sector of the 
linear sigma model does not feed back into the light quark sector, results using different mixing 
matrices may be used as a self-consistency check of the model in comparison with corresponding experimental
data. This will be discussed in more detail in section \ref{secIII}. 
In particular we use the mixing matrix $U_2$ from Ref.~\cite{Close:2005vf} and $U_{3..5}$ from Ref.~\cite{Zhao:2007ze}
\begin{equation}\label{U2}
U_2=\left(%
\begin{array}{ccccc}
 0.36 & 0.93 & 0.09\\
-0.84 & 0.35  &-0.41\\
 0.40 & -0.07 &-0.91\\
\end{array}%
\right)\,,
\end{equation}
\begin{equation}\label{U3}
U_3=\left(%
\begin{array}{ccccc}
 0.859 & 0.302 & 0.413\\
-0.128 & 0.908  & -0.399\\
 -0.495 & 0.290 & 0.819\\
\end{array}%
\right)\,, \vspace*{1cm}
U_4=\left(%
\begin{array}{ccccc}
 -0.06 & 0.97 & -0.24\\
0.89 & -0.06  &-0.45\\
 0.45 & 0.24 & 0.86\\
\end{array}%
\right)\,, \vspace*{1cm}
U_5=\left(%
\begin{array}{ccccc}
 -0.68 & 0.67 & -0.30\\
0.49 & 0.72  &-0.49\\
0.54 & 0.19 & 0.81\\
\end{array}%
\right).
\end{equation}

\section{Parameters and results}\label{secIII}

All the parameters and wave-function renormalization constants of the Lagrangian (\ref{Lagc}) have been fixed in Refs.~\cite{Parganlija:2012fy,Eshraim:2014eka}. 
Their values are summarized in Table \ref{Tab:ren}.

\begin{table}[H]
\centering
\begin{tabular}
[c]{|c|c|c|c|c|c|} \hline parameter & value  & parameter& value& renormalization factor & value \\
\hline $m_{1}^{2}$  & $0.413\times10^{6}$ MeV$^{2}$ &$\omega_{a1}$      & 0.00068384  & $Z_\pi=Z_{\eta_N}$ & 1.70927\\
\hline $m_{0}^{2}$  & $-0.918\times 10^{6}$ MeV$^{2}$ &$\omega_{f_{1N}}$ & 0.00068384 &  $Z_K$  & 1.60406 \\
\hline $\delta_{S}$ & $0.151\times10^{6}$MeV$^{2}$& $\omega_{f_{1S}}$ & 0.0005538 &  $Z_{\eta_C}$ & 1.11892   \\
\hline $\delta_{C}$ & $3.91\times10^{6}$MeV$^{2}$  & $\omega_{K1}$ & 0.000609921    &  $Z_{D_S}$   & 1.15716\\
\hline $\varepsilon_{C}$ & $2.23\times10^{6}$MeV$^{2}$&$\omega_{K^\ast}$ & -0.0000523i&  $Z_{D^\ast_0}$ & 1.00649  \\
\hline $c$   &  $1.987\times10^{-8}$MeV$^{-4} $  &$\omega_{D_{S1}}$ & 0.000203&  $Z_{D^{\ast0}_0}$ &1.00649\\
\hline  $g_{1}$ & $5.84$ &$\omega_{D^\ast}=\omega_{D^{\ast0}}$ & -0.0000523i & $Z_{\eta_S}$ & 1.53854 \\
\hline $h_{2}$ & $9.88$& $\omega_{D^\ast_0}$ & -0.0000423i & $Z_{K_{S}}$ & 1.00105 \\
\hline $\lambda_{2}$ & $68.3$ & $\omega_{D_1}$       & 0.00020& $Z_{D}$ & 1.15256\\
\hline $h_{3}$ & $3.87$ & $\omega_{\chi_{C1}}$  & 0.000138 & $Z_{D_{S0}^\ast}$ & 1.00437\\
\hline
\end{tabular}
\caption{ Parameters and wave-function renormalization constants.}
 \label{Tab:ren}
\end{table}
The wave-function renormalization constants for $\pi$ and $\eta_N$ are equal because of isospin symmetry, 
similar for $D^\ast_0$ and $D^{\ast0}$. The gluon condensate $G_0$ is equal to $\Lambda\approx 3.3$ GeV \cite{Janowski:2014ppa} 
in pure YM theory, which is also used in the present discussion.

The parameters $\lambda_1^S,\lambda_1^C$ and $h_1^S,h_1^C$ have either been set to zero (or not present at all)
in previous studies
\cite{Parganlija:2010fz,Eshraim:2014eka,Eshraim:2015cia} in agreement with large-${N_c}$ expectations.
In the latter works, masses of charmed mesons and the (OZI-dominant) strong decays of open charmed mesons have 
been considered, but not OZI-suppressed decays. However, as we will explain in the following, in these cases
small but non-zero values of $\lambda_1^C$ and $h_1^C$ are mandatory. 

\begin{figure}[b]
\begin{center}
\includegraphics[height=1in, width=2.3in]{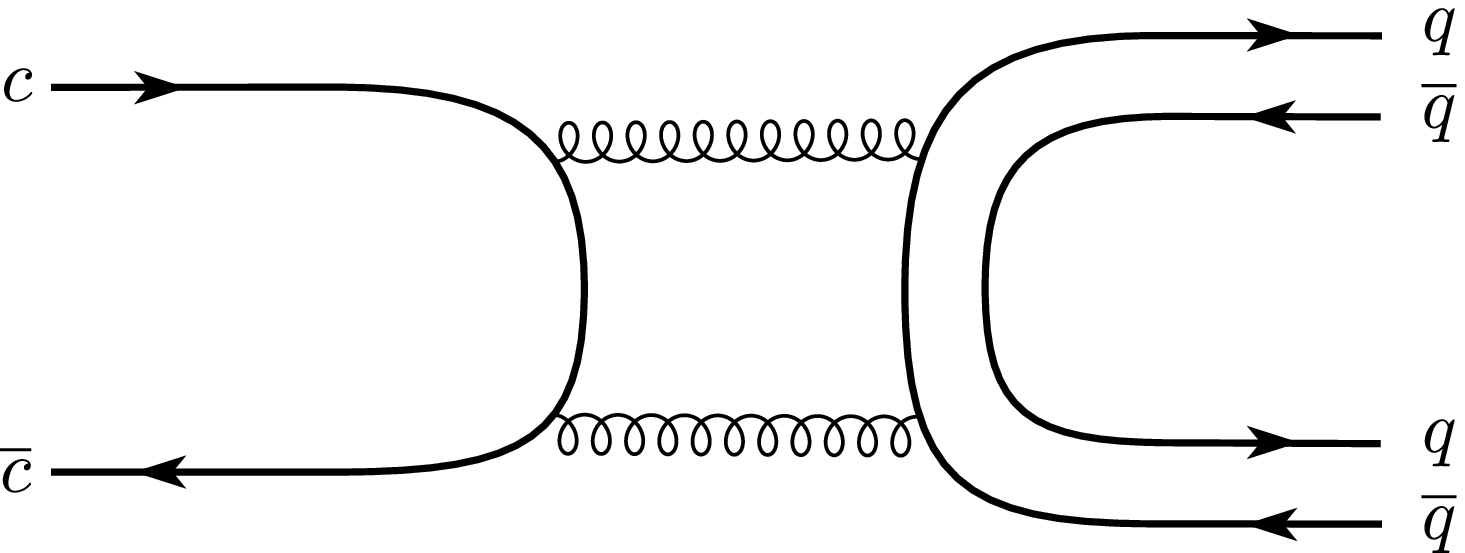}
\end{center}
\begin{center}
\includegraphics[height=1in, width=2.3in]{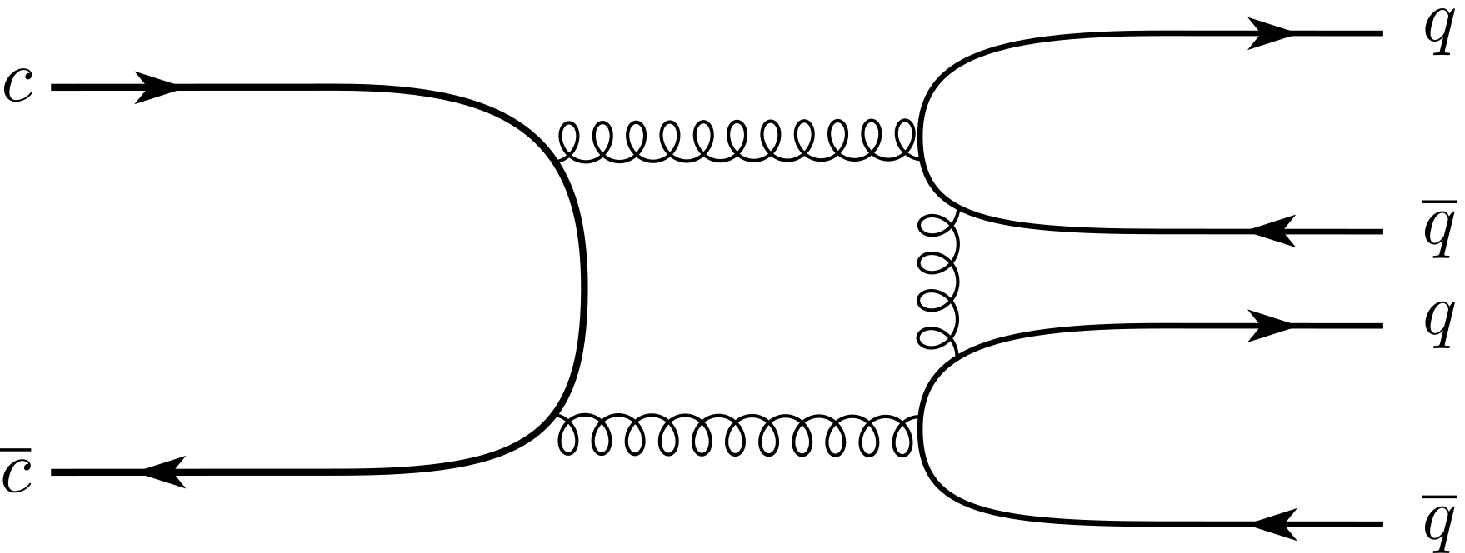}
\caption{Decay of charmonium state into two mesons. $q$ refers to the up (u), down (d), and strange (s) quark 
flavours.}
\label{figAVP}
\end{center}
\end{figure}

The decay of the charmonium states $\eta_c$ and $\chi_{c0}$ into hadrons is mediated by gluon annihilation. 
For the two-body decays, this annihilation can occur via two-gluon exchange diagrams for all contributions and
additional so-called double OZI-suppressed three-gluon exchange diagrams for decays into pairs of iso-singlet states,
cf. Fig.~\ref{figAVP}. Similar diagrams contribute to the three-particle decays. In these decays, gluons carry 
all the energy. Therefore, the interaction is relatively weak due to asymptotic freedom, which leads to 
OZI-suppression. As a consequence, the decay of the charmonium states $\eta_c$ and $\chi_{C0}$ into
(axial-)vector and (pseudo)scalar mesons or scalar glueballs is dynamically suppressed. In the eLSM, 
this is reflected by small but non-zero values of the large-$N_c$ suppressed parameters $\lambda_1^C$ and $h_1^C$.

We determine their size using experimental decay widths of $\chi_{C0}$ listed by the PDG
\cite{Patrignani:2016xqp} via a $\chi^2$ fit. To this end we employ the five known decay widths into
non-isosinglet final states given in table~\ref{Tab:chi1}. Using   
\begin{equation}
\chi^{2}(\lambda_1^C, h_1^C)\equiv\sum_{i}^{6}\bigg(\frac{\Gamma_{i}^{th}-\Gamma_{i}^{exp}}
{\Gamma_{i,error}^{exp}}\bigg)^{2}\text{ ,} \label{chi2,1}%
\end{equation}
we obtain 
\begin{equation}
\lambda_1^C=-0.161\pm0.005\,\,\,\text{and}\,\,\, h_1^C=0.046\pm0.003\,,\label{lam}
\end{equation}
with a reasonable $\chi^2/d.o.f= 3.8$. The values of the parameters $\lambda_1^C$ and $h_1^C$ are indeed small. 
A posteriori we therefore justify the assumptions of Refs.~\cite{Eshraim:2014eka, Eshraim:2015cia} concerning
the heavy quark sector. 
Reviewing the fit results, it is apparent that the decay into the iso-singlet $\phi$-mesons 
is not very well reproduced by the model (albeit with a large error). It dominates by far the $\chi^2$-value. 
Since this decay is distinguished from the others by the additional contributions from double OZI-suppressed 
diagrams (cf. Fig.~\ref{figAVP}), we conclude that these may not be well represented by the current model setup.
One could proceed by excluding the $\phi$-meson decay from the fitting procedure and strictly restrict the 
whole approach (in the current setup) to the non-iso-singlet sector. However, since we are particularly interested
in decays to scalar isoscalar states, we decided to keep the information from the $\phi$-meson decay in the fit
and accept a large error margin in our predictions for these channels. As a result we may expect (semi-)quantitative
predictions in the non-iso-singlet sector (where the fits works excellent), whereas our results for the 
iso-singlet decays should be taken as order of magnitude estimates. Certainly, this situation should be 
improved in the future.

\begin{table}[H]
\centering
\begin{tabular}
[c]{|c|c|c|c|} \hline Decay Channel &  theoretical result [MeV] &
Experimental result [MeV]\\
\hline $\Gamma_{\chi_{c0}\rightarrow \bar{K}^{*0}_0K^{*0}_0}$   & 0.010$\pm$0.003  & 0.010 
$\substack{+0.004\\-0.003}$ \\
\hline $\Gamma_{\chi_{c0}\rightarrow  K^-K^+}$ 					& 0.059$\pm$0.008  & 0.062$\pm$0.005\\
\hline $\Gamma_{\chi_{c0}\rightarrow \pi\pi}$      				& 0.090$\pm$0.011  & 0.087$\pm$0.006\\
\hline $\Gamma_{\chi_{c0}\rightarrow \bar{K}^{*0}K^{*0}}$ 		& 0.014$\pm$0.007  & 0.018$\pm$0.006\\
\hline $\Gamma_{\chi_{c0}\rightarrow \omega \omega}$    		& 0.012$\pm$0.006  & 0.010$\pm$0.001\\
\hline $\Gamma_{\chi_{c0}\rightarrow \phi \phi}$    			& 0.0035$\pm$0.0036& 0.0081$\pm$0.0009\\\hline
\hline $\Gamma_{\chi_{c0}\rightarrow \eta \eta}$ 				& 0.022$\pm$0.002  & 0.031$\pm$0.003 \\
\hline $\Gamma_{\chi_{c0}\rightarrow \eta' \eta'}$    			& 0.021$\pm$0.001  & 0.021$\pm$0.002\\
\hline $\Gamma_{\eta_{c}\rightarrow \eta \pi^{-} \pi^{+}}$   	& 0.12$\pm$0.02    &  0.54$\pm$0.16\\
\hline $\Gamma_{\eta_{c}\rightarrow \eta' \pi \pi}$  			& 0.081$\pm$0.019  &  1.30$\pm$0.54\\
\hline
\end{tabular}
\caption{The partial decay widths of $\chi_{c0}$ used to fix three model parameters. The upper six decays
fix $\lambda_1^C$ and $h_1^C$, whereas the lower four decays are used to fix $\delta \widetilde{c}$.}
 \label{Tab:chi1}
\end{table}

Furthermore, we adjust the coefficient $\delta \widetilde{c}$ of the 
modification of the axial anomaly term to fit the results of the decay widths of $\chi_{c0}$ 
into flavour singlet mesons. Here we use the 
decay widths of $\chi_{c0}$ into the experimentally known $\eta$ and $\eta'$-channels given in table~\ref{Tab:chi1}.
We perform a fit by minimizing the $\chi^2$-function,
\begin{align}
\chi^{2}(c_D)\equiv&\bigg(\frac{\Gamma_{\chi_{c0}\rightarrow \eta\eta}^{th}(c_D)-\Gamma_{\chi_{c0}\rightarrow \eta\eta}^{exp}}{\Gamma_{error}%
^{exp}}\bigg)^{2}+\bigg(\frac{\Gamma_{\chi_{c0}\rightarrow \eta'\eta'}^{th}(c_D)-\Gamma_{\chi_{c0}\rightarrow \eta'\eta'}^{exp}}{\Gamma_{error}%
^{exp}}\bigg)^{2}\\\nonumber%
&+\bigg(\frac{\Gamma_{\eta_{C}\rightarrow \eta'\pi\pi}^{th}(c_D)-\Gamma_{\eta_{C}\rightarrow \eta'\pi\pi}^{exp}}{\Gamma_{error}^{exp}}\bigg)^{2}+\bigg(\frac{\Gamma_{\eta_{C}\rightarrow \eta \pi^+ \pi^-}^{th}(c_D)-\Gamma_{\eta_{C}\rightarrow \eta \pi^+ \pi^-}^{exp}}{\Gamma_{error}%
^{exp}}\bigg)^{2}\text{ ,} \label{chi2}%
\end{align}
where $c_D=(c-\frac{1}{2}\phi_C \delta \widetilde{c})$. We obtain  
\begin{equation}
c_D=(7.255\pm0.0001)\times 10^{-10} \mbox{MeV}^{-4}
\end{equation}
with a $\chi^2/{\rm d.o.f}=5.4$. This value is equally dominated from the decays into two 
$\eta$ mesons and the two three-body decays. Also this result indicates that we may only expect 
order of magnitude estimates in the channels involving iso-singlets. 

\begin{table}[H]
\centering
\vspace*{3mm}
\begin{tabular}
[c]{|c|c|c|c|} \hline Decay Channel &  theoretical result [MeV] &
Experimental result [MeV]\\
\hline $\Gamma_{\chi_{c0}\rightarrow a_0 a_0}$  			& 0.0036$\pm$0.0019  	&   -\\
\hline $\Gamma_{\chi_{c0}\rightarrow K^{*} \overline{K}^{*}_0}$ & 0.000069$\pm$0.000049 	&   - \\
\hline $\Gamma_{\chi_{c0}\rightarrow \rho\rho}$  		& 0.010$\pm$0.006  	&   -\\
\hline $\Gamma_{\chi_{c0}\rightarrow \eta \eta'}$       & 0.0012$\pm$0.0005 & $<$0.0024 \\
\hline $\Gamma_{\chi_{c0}\rightarrow K^*_0 K \eta}$   	& 0.00042$\pm$0.00015  	& - \\
\hline $\Gamma_{\chi_{c0}\rightarrow K^*_0 K \eta'}$ 	& 0.00021$\pm$0.00013  	& - \\
\hline
\end{tabular}\\
\caption{The partial decay widths of $\chi_{c0}$ predicted by the model.}\label{Tab:chi2}
%\end{table}
%\begin{table}[t]
\centering
\vspace*{3mm}
\begin{tabular}
[c]{|c|c|c|c|c|c|c|c|} \hline Decay Channel &  $U_1$ (Ref.~\cite{Janowski:2014ppa})  
& $U_2$ (Ref.~\cite{Close:2005vf}) & $U_3$ (Ref.~\cite{Zhao:2007ze})& $U_4$ (Ref.~\cite{Zhao:2007ze}) 
& $U_5$ (Ref.~\cite{Zhao:2007ze}) &
Experimental result \cite{Patrignani:2016xqp}\\
\hline $\Gamma_{\chi_{c0}\rightarrow f_0(1370)f_0(1370)}$   &$ 5\times10^{-3}$  &$ 5\times10^{-3}$ &$ 4\times10^{-3}$   &$5\times10^{-3}$ &$ 1\times10^{-2}$   & $<3\times10^{-3}$\\
\hline $\Gamma_{\chi_{c0}\rightarrow f_0(1500)f_0(1500)}$   &$ 4\times10^{-3}$  &$ 2\times10^{-3}$ &$4\times10^{-3}$    &$2\times10^{-3}$    &$ 3\times10^{-3}$   & $<5\times10^{-4}$\\
\hline $\Gamma_{\chi_{c0}\rightarrow f_0(1370)f_0(1500)}$   &$ 1\times10^{-5}$  &$ 2\times10^{-4}$   &$ 9\times10^{-6}$   &$ 4\times10^{-4}$ &$ 7\times10^{-4}$   & $<2\times10^{-3}$\\
\hline $\Gamma_{\chi_{c0}\rightarrow f_0(1370)f_0(1710)}$   &$ 2\times10^{-4}$&$ 3\times10^{-5}$ &$ 3\times10^{-4}$ &$ 1\times10^{-6}$   &$ 5\times10^{-3}$ &    $(6.9\substack{+3.7\\-2.4} ) \times 10^{-3}$\\
\hline $\Gamma_{\chi_{c0}\rightarrow f_0(1500)f_0(1710)}$   &$ 3\times10^{-5}$&$ 1\times10^{-4}$   &$ 2\times10^{-5}$   &$ 5\times10^{-6}$   &$ 1\times10^{-4}$   &  $<7\times10^{-4}$\\
\hline $\Gamma_{\chi_{c0}\rightarrow f_0(1370) \eta\eta}$   &$ 3\times10^{-6}$  &$ 3\times10^{-8}$   &$ 9\times10^{-7}$   &$ 5\times10^{-7}$   &$ 5\times10^{-6}$   & -\\
\hline $\Gamma_{\chi_{c0}\rightarrow f_0(1500) \eta\eta}$   &$ 1\times10^{-5}$&$ 3\times10^{-6}$   &$ 2\times10^{-5}$ &$ 9\times10^{-11}$&$ 5\times10^{-6}$   &   -\\
\hline $\Gamma_{\chi_{c0}\rightarrow f_0(1370) \eta' \eta'}$&$ 1\times10^{-5}$  &$ 2\times10^{-5}$   &$ 2\times10^{-5}$   &$ 2\times10^{-5}$   &$ 3\times10^{-5}$   &   -\\
\hline $\Gamma_{\chi_{c0}\rightarrow f_0(1370) \eta \eta'}$ &$ 4\times10^{-10}$ &$ 9\times10^{-6}$   &$ 2\times10^{-5}$   &$ 2\times10^{-6}$   &$ 7\times10^{-5}$   & - \\
\hline $\Gamma_{\chi_{c0}\rightarrow f_0(1500) \eta\eta'}$  &$ 6\times10^{-6}$  &$ 4\times10^{-6}$   &$ 4\times10^{-5}$   &$ 2\times10^{-6}$   &$ 6\times10^{-6}$   & - \\
\hline $\Gamma_{\chi_{c0}\rightarrow f_0(1710) \eta \eta}$  &$ 8\times10^{-7}$  &$ 8\times10^{-6}$   &$ 6\times10^{-7}$   &$ 1\times10^{-5}$   &$ 6\times10^{-6}$   & - \\
\hline $\Gamma_{\chi_{c0}\rightarrow f_0(1710) \eta \eta'}$ &$ 4\times10^{-7}$  &$ 1\times10^{-5}$   &$ 2\times10^{-6}$   &$ 1\times10^{-5}$ &$ 1\times10^{-5}$ & - \\
\hline
\end{tabular}\\
\caption{The partial decay widths of $\chi_{c0}$ into scalar isoscalar mesons in units of MeV. 
The corresponding mixing matrices in the scalar isoscalar sector are given in
 Eqs.(\ref{scalmix1})-(\ref{U3}).}\label{Tab:chi3}
\end{table}

Having fixed all model parameters we now discuss the predictions of the model starting with the two- and
three-body decays in table~\ref{Tab:chi2} (all relevant expressions for the calculations are presented 
in the Appendix). As discussed above we do expect very reasonable predictions for the channels not 
involving iso-singlet states, i.e. the first three entries in the table, whereas we regard the second
three entries as order of magnitude estimates. In this respect it is satisfactory to see that the 
decay into the $\eta \eta'$ pair is in agreement with the experimental bounds.  

The two- and three-body decays of the $\chi_{c0}$ into the scalar-isoscalar resonances $f_0(1370)$, 
$f_0(1500)$, $f_0(1710)$ are shown in table~\ref{Tab:chi3}. Considering the intrinsic uncertainties of the model, 
we regard these results as order of magnitude estimates\footnote{Therefore we only give one significant digit
and refrain from giving the error due to the fitting procedure, since this error is at least an order 
of magnitude smaller than the systematic error due to model uncertainties.}. 
Nevertheless it is interesting to compare the 
results obtained with the mixing matrix of the eLSM, $U_1$ \cite{Janowski:2014ppa}, with the ones obtained
by other mixing patters. Compared with the experimental bounds we find that none of the mixing scenarios
are in agreement with all experimental bounds. However, considering the current experimental and theoretical 
uncertainties this statement is not very rigorous. Comparing the five different 
scenarios we find that in some channels the deviations are within an order of magnitude, whereas in other 
channels like $\Gamma_{\chi_{c0}\rightarrow f_0(1370) \eta\eta}$ or 
$\Gamma_{\chi_{c0}\rightarrow f_0(1370) \eta \eta'}$ differences of two or more magnitudes arise which may
be resolved by future experiments. Thus, in general, decays of charmonia offer the interesting possibility 
to involve the heavy quark sector to further constrain and explore glueball physics in the light quark sector.

\begin{table}[H]
\centering
\begin{tabular}
[c]{|c|c|c|c|} \hline Decay Channel &  Theoretical result [MeV] &
Experimental result [MeV]\\
\hline $\Gamma_{\eta_{c}\rightarrow\overline{K}^{*}_0K}$   	&0.010$\pm$0.006  	& - \\
\hline $\Gamma_{\eta_{c}\rightarrow a_0\pi}$      			&0.010 $\pm$0.007  	& - \\
\hline $\Gamma_{\eta_{c}\rightarrow \eta \eta \eta}$   		& 0.054$\pm$0.015  	& - \\
\hline $\Gamma_{\eta_{c}\rightarrow \eta' \eta' \eta'}$     & 0.0024$\pm$0.0032  	& - \\
\hline $\Gamma_{\eta_{c}\rightarrow \eta'\eta \eta}$ 		& 0.045$\pm$0.014  	& - \\
\hline $\Gamma_{\eta_{c}\rightarrow \eta' \eta' \eta}$    	& 0.0036$\pm$0.0039  	& - \\
\hline $\Gamma_{\eta_{c}\rightarrow \eta K \overline{K}}$   & 0.16$\pm$0.03  	& 0.43$\pm$0.05\\
\hline $\Gamma_{\eta_{c}\rightarrow \eta' K K}$      		& 0.43$\pm$0.04 	& - \\
\hline $\Gamma_{\eta_{c}\rightarrow K K \pi}$     			& 0.096$\pm$0.020  	& - \\
\hline
\end{tabular}\\
\caption{The partial decay widths of $\eta_{c}$ into two and three mesons.}\label{eta1}
%\end{table}
%\begin{table}[b]
%\centering
\vspace*{3mm}
\begin{tabular}
[c]{|c|c|c|c|c|c|c|c|} \hline Decay Channel &   $U_1$ (Ref.~\cite{Janowski:2014ppa})  
& $U_2$ (Ref.~\cite{Close:2005vf}) & $U_3$ (Ref.~\cite{Zhao:2007ze})& $U_4$ (Ref.~\cite{Zhao:2007ze}) 
& $U_5$ (Ref.~\cite{Zhao:2007ze}) &
Experimental result \cite{Patrignani:2016xqp}\\
\hline $\Gamma_{\eta_{c}\rightarrow f_0(1370) \eta}$  &$ 1\times10^{-3}$ &$ 5\times10^{-3}$ 	&$ 9\times10^{-3}$ 	&$ 8\times10^{-3}$   &$ 2\times10^{-2}$   &- \\
\hline $\Gamma_{\eta_{c}\rightarrow f_0(1500) \eta}$  &$ 2\times10^{-2}$ 	&$ 4\times10^{-4}$  &$ 8\times10^{-3}$ 	&$ 1\times10^{-3}$ 	&$ 4\times10^{-4}$ & seen \\
\hline $\Gamma_{\eta_{c}\rightarrow f_0(1710) \eta}$  &$ 2\times10^{-4}$&$ 1\times10^{-2}$  	&$4\times10^{-3}$ 	&$ 1\times10^{-2}$   	&$1\times10^{-2}$ 	& - \\
\hline $\Gamma_{\eta_{c}\rightarrow f_0(1370) \eta'}$ &$ 2\times10^{-1}$ 	&$ 2\times10^{-1}$  	&$ 2\times10^{-1}$ 	&$ 2\times10^{-1}$   	&$ 4\times10^{-1}$   	& - \\
\hline $\Gamma_{\eta_{c}\rightarrow f_0(1500) \eta'}$ &$ 1\times10^{-1}$ 	&$ 1\times10^{-2}$ 	&$ 1\times10^{-5}$ &$ 5\times10^{-2}$   	&$ 3\times10^{-2}$   	&- \\
\hline $\Gamma_{\eta_{c}\rightarrow f_0(1710) \eta'}$ &$ 4\times10^{-3}$ &$ 4\times10^{-2}$  	&$ 5\times10^{-2}$ 	&$ 6\times10^{-3}$   &$ 9\times10^{-2}$ 	& - \\
\hline
\end{tabular}\\
\caption{The partial decay widths of $\eta_{c}$ into scalar isoscalar states in units of MeV.}\label{eta2}
\end{table}

Next we discuss the decay widths of the pseudoscalar charmonium state $\eta_C (1P)$ into (pseudo)scalar 
mesons given in tables \ref{eta1}. From the results of the fitting procedure again we may expect 
predictions only on the order of magnitude level. Indeed, this is satisfied in the only channel 
where experimental data are available, $\eta_c \rightarrow \eta K \bar{K}$.
For the decays of the $\eta_c$ into the isoscalar scalar and pseudoscalar states shown in table \ref{eta2}
we find again a high sensitivity to the mixing matrix in some channels, whereas others are much less sensitive.
Again, this result may provide guidance in the data analysis of future experiments on an order of magnitude basis. 

Finally, let us discuss the decay of $\eta_c$ into a pseudoscalar glueball $\widetilde{G}$. It proceeds via 
the channel $\eta_C\rightarrow \pi\pi\widetilde{G}$. The width $\Gamma_{\eta_C\rightarrow \pi\pi\widetilde{G}}$ 
depends on the coupling constant $c_{\widetilde{G}\Phi}$ which can be determined
by the following relation
\begin{equation}
c_{\widetilde{G}\Phi}=\frac{\sqrt{2}\,c_{\widetilde{G}\Phi(N_f=3)}}{\phi_C}\,.\label{cGphi}%
\end{equation}
where $c_{\widetilde{G}\Phi(N_f=3)}= 4.48\pm 0.46$ has been determined in Ref.~\cite{Eshraim:2012jv} via 
the decay of the pseudoscalar glueball into scalar and pseudoscalar mesons. 
We then obtain $c_{\widetilde{G}\Phi}=0.036$ in the present four-flavour case. 
In order to determine the decay width, the mass of the resulting pseudoscalar glueball is important. 
In the literature, sometimes the state $\eta(1405)$ has been considered as a candidate for a light 
pseudoscalar glueball, see e.g. \cite{Gutsche:2009jh}. However, this identification may be questionable,
in particular since the (quenched) lattice results point to much heavier masses 
\cite{Morningstar:1999rf,Chen:2005mg}. Here we consider two cases, $m_{\widetilde{G}} = 2.6$ GeV 
as found on the lattice and a putative lower mass of $m_{\widetilde{G}} = 2.37$ GeV 
as suggested e.g. by the identification of the pseudoscalar glueball with the resonance 
$X(2370)$. We then find 
\begin{equation}
\Gamma_{\eta_c\rightarrow \pi\pi\widetilde{G}(2600)}=0.124\,\, {\rm MeV},\,\,\,\,\,\,\, \Gamma_{\eta_c\rightarrow \pi\pi\widetilde{G}(2370)}=0.160\,\, {\rm MeV}\,, \label{etadpg}%
\end{equation}
showing a not too large variation of the decay width with glueball mass.

\section{Conclusion and outlook}\label{secIV}

In this work we have further extended the extended linear sigma model to be able to deal with the 
strong decays of the $\eta_c$ and the $\chi_{c0}$. Encouraged by the unexpected success of 
the model to deal with the spectrum of charmonium and open charm decays \cite{Eshraim:2014eka}, we
determined various two- and three-body decays into (pseudo-)scalar and vector mesons and a putative 
pseudoscalar glueball. Unfortunately, the
study is not as conclusive as one could wish for. We obtain excellent results for the non-iso-singlet
two-body decays of the $\chi_{c0}$, where we also presented predictions for some as yet unmeasured channels. 
The singlet decays, however, give mixed results, with some very reasonable but others only correct on the 
order of magnitude level. Consequently, the predictive power of the model in this sector remains on the 
order of magnitude level. Since, on the other hand, experimental results
are entirely missing for a large range of decays, we nevertheless consider these order of magnitude 
estimates to be a helpful guidance. Particularly interesting for future experimental and theoretical 
studies could be the decay channels into isoscalar scalar mesons, which are supposed to contain admixtures
from a scalar glueball. Different mixing matrices lead to a range of different decay patterns for 
the $\eta_c$ and the $\chi_{c0}$, which could be used to discriminate between different mixing scenarios.
To this end one needs to further refine the model and focus on these channels in ongoing and future 
experiments such as BESIII, Belle II, LHCb and the PANDA experiment at the FAIR facility.

\section*{Acknowledgments}

The authors thank F. Giacosa, F. Maas, D. H. Rischke and B.-J. Schaefer for useful discussions. 
We thank the referee for pointing out the need to treat the parameters $\lambda_1^{S,C}$ and $h_1^{S,C}$ 
different in the light and heavy quark sector.
This work was supported by the BMBF under contract No. 05H15RGKBA and the Helmholtz International Center 
for FAIR within the LOEWE program of the State of Hesse.

\appendix

\section{Decay widths}

The general formula for the two-body decay width is given by \cite{Patrignani:2016xqp}:
\begin{equation}
\label{B1}\Gamma_{A\rightarrow BC}=S_{A\rightarrow BC}\frac{k(m_{A}%
,\,m_{B},\,m_{C})}{8 \pi m_{A}^{2}}|\mathcal{M}_{A\rightarrow
BC}|^{2}\,.
\end{equation}
The center-of-mass momentum $k(m_{A},\,m_{B},\,m_{C})$ of the decay products B,C is given by
\begin{equation}
k(m_{A},\,m_{B},\,m_{C})=\frac{1}{2m_{A}}\sqrt{m_{A}^{4}+(m_{B}^{2}-m_{C}%
^{2})^{2}-2m_{A}^{2}\,(m_{B}^{2}+m_{C}^{2})}\theta(m_{A}-m_{B}-m_{C})\,.
\label{B2}%
\end{equation}
$\mathcal{M}_{A\rightarrow BC}$ is the corresponding tree-level decay 
amplitude, and $S_{A\rightarrow BC}$ denotes a symmetrization factor 
(it equals $1$ if B and C are different and it equals $1/2$ for two 
identical particles in the final state).

\begin{figure}[H]
\begin{center}
\includegraphics[width=5cm]{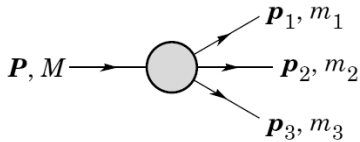}
\caption{Feynman diagram of the three body decay.} \label{fig:3bodydecay}
\end{center}
\end{figure}
Using the notation of Fig.\ref{fig:3bodydecay} and the definition $m_{ij}=m_i+m_j$ the corresponding expression 
for the three-body decay width of the process $A\rightarrow B_{1}B_{2}B_{3}$ reads \cite{Patrignani:2016xqp}:
\begin{align}
\Gamma_{A\rightarrow B_{1}B_{2}B_{3}}&=\frac{S_{A\rightarrow
B_{1}B_{2}B_{3}}}{32(2\pi)^{3}M^{3}}\int_{(m_{1}+m_{2})^{2}%
}^{(M-m_{3})^{2}}|-i\mathcal{M}_{A\rightarrow B_{1}B_{2}B_{3}}%
|^{2}\nonumber\\&\times\sqrt{\frac{(-m_{1}+m_{12}-m_{2})(m_{1}+m_{12}-m_{2})(-m_{1}%
+m_{12}+m_{2})(m_{1}+m_{12}+m_{2})}{m_{12}^{2}}}\nonumber\\&\times
\sqrt{\frac{(-M+m_{12}-m_{3})(M+m_{12}-m_{3}%
)(-M+m_{12}+m_{3})(M+m_{12}+m_{3})}{m_{12}^{2}}%
}dm_{12}^{2}\text{ ,}\label{3bodydecay}
\end{align}
where $\mathcal{M}_{A\rightarrow B_{1}B_{2}B_{3}}$ is the
corresponding tree-level decay amplitude, and $S_{A\rightarrow
B_{1}B_{2}B_{3}}$ is a symmetrization factor (equal to $1$ if the 
$B_{i}$ are all different, equal to $2$ for two identical particles 
in the final state, and equal to $6$ for three identical particles).

\section{Decay rates for $\chi_{c0}$ \label{Sec2}}

We present the explicit expressions for the two- and three-body decay
rates for the scalar hidden-charmed meson $\chi_{c0}$.

\subsection{Two-body decay rates for $\chi_{C0}$}

The explicit expressions for the two-body decay rates of $\chi_{c0}$ are extracted from the
Lagrangian (\ref{Lagc}) and are presented in the following.

\textbf{Decay channel $\chi_{C0}\rightarrow\overline{K}^{\ast0}_{0}K^{\ast0}_{0}$ }\\

The corresponding interaction part of the Lagrangian (\ref{Lagc}) reads
\begin{align}
\mathcal{L}_{\chi_{C0}\overline{K}^{\ast}_{0}K^{\ast}_{0}}=&-2\lambda_1^C
\,
Z_{K^\ast_0}^2\,\phi_C\,\chi_{C0}(\overline{K}^{\ast0}_{0}K^{\ast0}_{0}+
K^{\ast-}_{0}K^{\ast+}_{0})\nonumber\\&-h_1^C\,\phi_C\,Z_{K^\ast_0}^2
\omega^2_{K^\ast} \chi_{C0}(\partial_\mu K^{\ast0}_0\partial^\mu
\overline{K}^{\ast0}_0 + \partial_\mu K^{\ast-}_0\partial^\mu
K^{\ast+}_0 )\,.
\end{align}
Consider only the
$\chi_{C0}\rightarrow\overline{K}^{\ast0}_{0}K^{\ast0}_{0}$ decay
channel, the $\chi_{C0}\rightarrow
K^{\ast-}_{0}K^{\ast+}_{0}$ will give the same contribution due to isospin symmetry,
\begin{align}
\mathcal{L}_{\chi_{C0}\overline{K}^{\ast0}_{0}K^{\ast0}_{0}}=&-2\lambda_1^C
\,
Z_{K^\ast_0}^2\,\phi_C\,\chi_{C0}\overline{K}^{\ast0}_{0}K^{\ast0}_{0}-h_1^C\,\phi_C\,Z_{K^\ast_0}^2
\omega^2_{K^\ast} \chi_{C0}\partial_\mu K^{\ast0}_0\partial^\mu
\overline{K}^{\ast0}_0 \,.
\end{align}
Let us denote the momenta of ${K}^{\ast0}_{0}$ and
$\overline{K}^{\ast0}_{0}$ as $P_1$ and $P_2$, respectively. The
energy-momentum conservation on the vertex implies $P=P_1+P_2$, where $P$
denotes the momenta of the decaying particle $\chi_{C0}$. Given
that our particles are on-shell, we obtain
\begin{equation}
P_1 \cdot P_2=\frac{P^2-P_1^2-P_2^2}{2}=\frac{m_{\chi_{C0}}^2-2\,m_{K^{\ast0}_0}}{2}\,.
\end{equation}
Upon substituting $\partial_\mu\rightarrow -iP^\mu$ for the decay
particle and $\partial_\mu\rightarrow +iP_{1,2}^\mu$ for the outgoing
particles, one obtains
\begin{align}
\mathcal{L}_{\chi_{C0}\overline{K}^{\ast0}_{0}K^{\ast0}_{0}}=\phi_C\,Z_{K^\ast_0}^2\bigg[-2\lambda_1^C+h_1^C\omega^2_{K^\ast}\frac{m_{\chi_{c0}}^2-2m_{K^\ast_0}^2}{2}\bigg]\chi_{C0}
K^{\ast0}_0\overline{K}^{\ast0}_0\,.
\end{align}
Consequently, the decay amplitude is given by
\begin{equation}
-i\mathcal{M}_{\chi_{C0}\rightarrow\overline{K}^{\ast0}_{0}K^{\ast0}_{0}}=i
\phi_C\,Z_{K^\ast_0}^2\bigg[2\lambda_1^C-h_1^C\omega^2_{K^\ast} \frac{m_{\chi_{c0}}^2-2m_{K^\ast_0}^2}{2}\bigg]\,.
\end{equation}
The decay width is obtained as
\begin{equation}
\Gamma_{\chi_{C0}\rightarrow\overline{K}^{\ast0}_{0}K^{\ast0}_{0}}=\frac{|\overrightarrow{k}_1|}{8\pi\,m_{\chi_{C0}}^2}|-i\mathcal{M}_{\chi_{C0}\rightarrow\overline{K}^{\ast0}_{0}K^{\ast0}_{0}}|^2\,.
\end{equation}
%where
%\begin{equation}
%|\overrightarrow{k}_1|=\frac{1}{2m_{\chi_{C0}}}\bigg[m_{\chi_{C0}}^4+(m_{\overline{K}^{\ast0}_{0}}^2-m_{K^{\ast0}_{0}}^2)^2-2(m_{\overline{K}^{\ast0}_{0}}^2+m_{K^{\ast0}_{0}}^2)m_{\chi_{C0}}^2\bigg]^{1/2}\,.
%\end{equation}

\vspace*{5mm}
\textbf{Decay channel $\chi_{C0}\rightarrow K^-K^+$ }\\

The corresponding interaction part of the Lagrangian (\ref{Lagc}) reads
\begin{align}
\mathcal{L}_{\chi_{C0}KK}=&-2\lambda_1^C \,
Z_{K}^2\,\phi_C\,\chi_{C0}(\overline{K}^{0}K^{0}+
K^{-}K^{+}_{0})\nonumber\\&+h_1^C\,\phi_C\,Z_{K}^2 \omega^2_{K_1}
\chi_{C0}(\partial_\mu K^{0}\partial^\mu \overline{K}^{0} +
\partial_\mu K^{-}\partial^\mu K^{+})\,.
\end{align}
In a similar way as the previous case, one can obtain the decay
width for the channel $\chi_{C0}\rightarrow K^-K^+$ as
\begin{equation}
\Gamma_{\chi_{C0}\rightarrow K^-K^+
}=\frac{|\overrightarrow{k}_1|}{8\pi\,m_{\chi_{C0}}^2}\,\phi_C^2\,Z_{K}^4
\bigg[2\lambda_1^C+h_1^C\omega^2_{K_1}\bigg(\frac{m_{\chi_{c0}}^2-2m_{K}^2}{2}\bigg)\bigg]^2\,.
\end{equation}
%where
%\begin{equation}
%|\overrightarrow{k}_1|=\frac{1}{2m_{\chi_{C0}}}\bigg[m_{\chi_{C0}}^4-4m_{K}^2\,m_{\chi_{C0}}^2\bigg]^{1/2}\,.
%\end{equation}

\vspace*{5mm}
\textbf{Decay channel $\chi_{C0}\rightarrow\pi\pi$}\\

The corresponding interaction part of the Lagrangian (\ref{Lagc}) reads
\begin{align}
\mathcal{L}_{\chi_{C0}\pi\pi}=-\lambda_1^C\,\phi_C\,
Z_{\pi}^2\,\chi_{C0}({\pi^0}^2+2
\pi^{-}\pi^{+})+\frac{1}{2}h_1^C\,\phi_C\,Z_{\pi}^2 \omega^2_{a_1}
\chi_{C0}[(\partial_\mu\pi^0)^2+2\partial_\mu \pi^-\partial^\mu
\pi^+]\,,
\end{align}
and leads to the decay width 
\begin{equation}
\Gamma_{\chi_{C0}\rightarrow\pi\pi}=\frac{3}{2}\frac{|\overrightarrow{k}_1|}{8\pi\,m_{\chi_{C0}}^2}\,\phi_C^2\,Z_{\pi}^4
\bigg[2\lambda_1^C+h_1^C\omega^2_{a_1}\bigg(\frac{m_{\chi_{c0}}^2-2m_{\pi}^2}{2}\bigg)\bigg]^2\,.
\end{equation}
%with
%\begin{equation}
%|\overrightarrow{k}_1|=\frac{(m_{\chi_{C0}}^4-4m_\pi^2
%m_{\chi_{C0}}^2)^{1/2}}{2m_{\chi_{C0}}}\,.
%\end{equation}

\vspace*{5mm}
\textbf{Decay channel $\chi_{C0}\rightarrow\overline{K}^{\ast0}K^{\ast0}$}\\

The corresponding interaction Lagrangian is extracted as
\begin{align}
\mathcal{L}_{\chi_{C0}\overline{K}^{\ast0}K^{\ast0}}=h_1^C\,\phi_C\,\chi_{C0}(
K^{\ast-}_\mu
K^{\ast+\mu}+K^{\ast0}_\mu\overline{K}^{\ast0\mu})\,.
\end{align}
Consider only the $K^{\ast0}_\mu\overline{K}^{\ast0\mu}$ decay
channel, then
\begin{align}
\mathcal{L}_{\chi_{C0}\overline{K}^{\ast0}K^{\ast0}}=h_1^C\,\phi_C\,\chi_{C0}K^{\ast0}_\mu\overline{K}^{\ast0\mu}\,.
\end{align}
Put
\begin{equation}
 A_{\chi_{C0}\overline{K}^{\ast0}K^{\ast0}}= h_1^C\,\phi_C\,.
\end{equation}

Let us denote the momenta of $\chi_{C0}$, $\overline{K}^{\ast0}$,
and $K^{\ast0}$ as $P$, $P_{1}$, and $P_{2}$, respectively, while
the polarisation vectors are denoted as
$\varepsilon_{\mu}^{(\alpha)}(P_{1})$ and
$\varepsilon_{\nu}^{(\beta )}(P_{2})$. Then, upon substituting
$\partial^{\mu}\rightarrow iP_{1,2}^{\mu}$ for the outgoing
particles, we obtain the following Lorentz-invariant
$\chi_{C0}\overline{K}^{\ast0}K^{\ast0}$ scattering amplitude
$-i\mathcal{M}^{(\alpha,\beta)}_{\chi_{C0}\rightarrow\overline{K}^{\ast0}K^{\ast0}}$:\\

\begin{equation}
-i\mathcal{M}_{\chi_{C0}\rightarrow\overline{K}^{\ast0}K^{\ast0}}^{(\alpha,\beta)}=\varepsilon_{\mu
}^{(\alpha)}(P_{1})\varepsilon_{\nu}^{(\beta)}(P_{2})h_{\chi_{C0}\overline{K}^{\ast0}K^{\ast0}}^{\mu\nu
}\,,\label{iMSVV}
\end{equation}

with

\begin{equation}
h_{\chi_{C0}\overline{K}^{\ast0}K^{\ast0}}^{\mu\nu}=iA_{\chi_{C0}\overline{K}^{\ast0}K^{\ast0}}g^{\mu\nu}\text{,}
\label{hSVV}
\end{equation}
$$\\$$
where $h_{\chi_{C0}\overline{K}^{\ast0}K^{\ast0}}^{\mu\nu}$
denotes the $\chi_{C0}\overline{K}^{\ast0}K^{\ast0}$
vertex.\newline

The averaged squared amplitude $|\overline{-i\mathcal{M}}|^{2}$ is
determined as follows:

\begin{align}
\left\vert
\overline{-i\mathcal{M}_{\chi_{C0}\rightarrow\overline{K}^{\ast0}K^{\ast0}}}\right\vert
^{2}
=&\frac{1}{3}\sum\limits_{\alpha,\beta=1}^{3}\left\vert
-i\mathcal{M}_{\chi_{C0}\rightarrow\overline{K}^{\ast0}K^{\ast0}}^{(\alpha,\beta)}\right\vert
^{2}\nonumber\\
=&\frac{1}{3}\sum\limits_{\alpha,\beta=1}^{3}\varepsilon_{\mu}^{(\alpha
)}(P_{1})\varepsilon_{\nu}^{(\beta)}(P_{2})h_{\chi_{C0}\overline{K}^{\ast0}K^{\ast0}}^{\mu\nu}\varepsilon
_{\kappa}^{(\alpha)}(P_{1})\nonumber\\
&\,\,\,\,\,\,\,\,\,\,\,\,\,\,\,\,\,\,\,\,\,\times \varepsilon_{\lambda}^{(\beta)}(P_{2})h_{\chi_{C0}\overline{K}^{\ast0}K^{\ast0}
}^{\ast\kappa\lambda}\text{ .}\label{iMSVV1}
\end{align}

Then,

\begin{align}
|\overline{-i\mathcal{M}_{\chi_{C0}\rightarrow\overline{K}^{\ast0}K^{\ast0}}}|^{2}=\frac{1}{3}\bigg[
\left\vert
h_{\chi_{C0}\overline{K}^{\ast0}K^{\ast0}}^{\mu\nu}\right\vert
^{2}&-\frac{\left\vert h_{\chi_{C0}\overline{K}^{\ast0}K^{\ast0}}
^{\mu\nu}P_{1\mu}\right\vert ^{2}}{m_{V_{1}}^{2}}-\frac{\left\vert
h_{\chi_{C0}\overline{K}^{\ast0}K^{\ast0}
}^{\mu\nu}P_{2\nu}\right\vert
^{2}}{m_{V_{2}}^{2}}\nonumber\\
&+\frac{\left\vert
h_{\chi_{C0}\overline{K}^{\ast0}K^{\ast0}
}^{\mu\nu}P_{1\mu}P_{2\nu}\right\vert
^{2}}{m_{V_{1}}^{2}m_{V_{2}}^{2} }\bigg]  \text{ .}\label{iMSVV2}
\end{align}

From Eq.\ (\ref{hSVV}) we obtain 

$$h_{\chi_{C0}\overline{K}^{\ast0}K^{\ast0}}^{\mu\nu}P_{1\mu}=iA_{\chi_{C0}\overline{K}^{\ast0}K^{\ast0}%
}P_{1}^{\nu}\,,$$
$$h_{\chi_{C0}\overline{K}^{\ast0}K^{\ast0}}^{\mu\nu}P_{2\nu}=iA_{\chi_{C0}\overline{K}^{\ast0}K^{\ast0}}P_{2}^{\mu}\,,$$
and
$$h_{\chi_{C0}\overline{K}^{\ast0}K^{\ast0}}^{\mu\nu}P_{1\mu}P_{2\nu}=iA_{\chi_{C0}\overline{K}^{\ast0}K^{\ast0}}P_{1}\cdot P_{2}\,,$$ and
consequently

\begin{equation}
|\overline{-i\mathcal{M}_{\chi_{C0}\rightarrow\overline{K}^{\ast0}K^{\ast0}}}|^{2}=\frac{1}{3}\left[
4-\frac{P_{1}^{2}}{m_{K^{\ast0}}^{2}}-\frac{P_{2}^{2}}{m_{\overline{K}^{\ast0}}^{2}}+\frac
{(P_{1}\cdot
P_{2})^{2}}{m_{K^{\ast0}}^{2}m_{\overline{K}^{\ast0}}^{2}}\right]
A_{\chi_{C0}\overline{K}^{\ast0}K^{\ast0}}
^{2}\text{.}\label{iMSVV3}
\end{equation}

For on-shell states,
$P_{1,2}^{2}=m_{\overline{K}^{\ast0},K^{\ast0}}^{2}$ and Eq.\
(\ref{iMSVV3}) reduces to

\begin{align}
|\overline{-i\mathcal{M}_{\chi_{C0}\rightarrow\overline{K}^{\ast0}K^{\ast0}}}|^{2}& =\frac{1}{3}\left[
2+\frac{(P_{1}\cdot
P_{2})^{2}}{m_{\overline{K}^{\ast0}}^{2}m_{K^{\ast0}}^{2}}\right]
A_{\chi_{C0}\overline{K}^{\ast0}K^{\ast0} }^{2}\nonumber\\
&=\frac{1}{3}\left[
2+\frac{(m_{\chi_{C0}}^{2}-m_{\overline{K}^{\ast0}}^{2}-m_{K^{\ast0}}^{2})^{2}
}{4m_{\overline{K}^{\ast0}}^{2}m_{K^{\ast0}}^{2}}\right]
A_{\chi_{C0}\overline{K}^{\ast0}K^{\ast0}}^{2}\text{.}\label{iMSVV4}
\end{align}

Consequently, the decay width is

\begin{equation}
\Gamma_{\chi_{C0}\rightarrow\overline{K}^{\ast0}K^{\ast0}}=\frac{|\overrightarrow{k}_1|}{8\pi
m_{\chi_{C0}}^{2}}|\overline{-i\mathcal{M}_{\chi_{C0}\rightarrow\overline{K}^{\ast0}K^{\ast0}}}|^{2}\,.\label{GSVV}
\end{equation}
%where
%\begin{equation}
%|\overrightarrow{k}_1|=\frac{1}{2m_{\chi_{C0}}}[m_{\chi_{C0}}^4+(m^2_{\overline{K}^{\ast0}}-m^2_{{K}^{\ast0}})^2-2(m^2_{\overline{K}^{\ast0}}+m^2_{{K}^{\ast0}})^2 m_{\chi_{C0}}^2]^{1/2}\,.
%\end{equation}

\vspace*{5mm}
\textbf{Decay channel $\chi_{C0}\rightarrow \omega\omega$}\\

The corresponding interaction Lagrangian is extracted as
\begin{align}
\mathcal{L}_{\chi_{C0}\omega\omega}=\frac{1}{2}h_1^C\,\phi_C\,\chi_{C0}\omega_N^\mu\omega_{N\mu}\,,
\end{align}
which also has the same form as the interaction Lagrangian
$\mathcal{L}_{\chi_{C0}\rightarrow\overline{K}^{\ast0}K^{\ast0}}$. Thus one can obtain the decay width of $\chi_{C0}\rightarrow
\omega\omega$ as
\begin{equation}
\Gamma_{\chi_{C0}\rightarrow
\omega\omega}=2\frac{[m_{\chi_{C0}}^{4}-4m_\omega^2m_{\chi_{C0}}^{2}]^{1/2}}{16\pi m_{\chi_{C0}}^{3}}\,\, \frac{1}{4}(h_1^C)^2\,\phi_C^2\bigg[2+\frac{(m_{\chi_{C0}}^2-2m_{\omega}^2)^2}{4m_{\omega}^4}\bigg]\,.\label{dw.1}
\end{equation}

\vspace*{5mm}
\textbf{Decay channel $\chi_{C0}\rightarrow \phi\phi$}\\

The corresponding interaction Lagrangian is extracted as
\begin{align}
\mathcal{L}_{\chi_{C0}
\omega\omega}=\frac{1}{2}h_1^C\,\phi_C\,\chi_{C0}\omega_S^\mu\omega_{S\mu}\,.
\end{align}
Similar to the decay width for $\chi_{C0}\rightarrow
\omega\omega$, the decay width for $\chi_{C0}\rightarrow \phi\phi$
is
\begin{equation}
\Gamma_{\chi_{C0}\rightarrow
\phi\phi}=2\frac{[m_{\chi_{C0}}^{4}-4m_\phi^2m_{\chi_{C0}}^{2}]^{1/2}}{16 \pi m_{\chi_{C0}}^{3}}\,\, \frac{1}{4}(h_1^C)^2\,\phi_C^2\bigg[2+\frac{(m_{\chi_{C0}}^2-2m_{\phi}^2)^2}{4m_{\phi}^4}\bigg]\,.\label{dw.2}
\end{equation}

\vspace*{5mm}
\textbf{Decay channel $\chi_{C0}\rightarrow \rho\rho$}\\

The corresponding interaction Lagrangian is extracted as
\begin{align}
\mathcal{L}_{\chi_{C0}\rho\rho}=\frac{1}{2}h_1^C\,\phi_C\,\chi_{C0}(\rho^{0\mu}\rho^0_\mu+2\rho^{-\mu}\rho^+_\mu)\,,
\end{align}
which also has the same form as
$\mathcal{L}_{\chi_{C0}\omega\omega}$. We thus obtain the decay
width as
\begin{align}
\Gamma_{\chi_{C0}\rightarrow\rho\rho}&=3\Gamma_{\chi_{C0}\rightarrow\rho^0\rho^0}\nonumber\\&
=3\frac{[m_{\chi_{C0}}^{4}-4m_{\rho^0}^2m_{\chi_{C0}}^{2}]^{1/2}}{16\pi m_{\chi_{C0}}^{3}}\times \frac{1}{12}(h_1^C)^2\,\phi_C^2\bigg[2+\frac{(m_{\chi_{C0}}^2-2m_{\rho^0}^2)^2}{4m_{\rho^0}^4}\bigg]\,.\label{dw.3}
\end{align}

\vspace*{5mm}
\textbf{Decay channel $\chi_{C0}\rightarrow a_0 a_0$}\\

The corresponding interaction Lagrangian has the form
\begin{align}
\mathcal{L}_{\chi_{C0}a_0a_0}=&-\lambda_1^C
\,\phi_C\,\chi_{C0}({a_0^0}^2+2a_0^-\,a_0^+)\,.
\end{align}
The decay width of $\chi_{C0}$ into $a_0a_0$ can be obtained as
\begin{align}
\Gamma_{\chi_{C0}\rightarrow
a_0\,a_0}&=3\Gamma_{\chi_{C0}\rightarrow a^0_0\,a^0_0}\nonumber\\&
=3\frac{(\lambda_1^C)^2\,\phi_C^2}{8\pi\,m_{\chi_{C0}}^2}\,\,\frac{[m_{\chi_{C0}}^4-4m_{a_0^0}^2\,m_{\chi_{C0}}^2]^{1/2}}{2m_{\chi_{C0}}}\,.
\end{align}

\vspace*{5mm}
\textbf{Decay channel $\chi_{C0}\rightarrow K^{\ast} \overline{K}^{\ast}_{0} $}\\

The corresponding interaction Lagrangian from the Lagrangian
(\ref{Lagc}) reads
\begin{align}
\mathcal{L}_{\chi_{C0}K^{\ast0}
\overline{K}^{\ast0}_{0}}=Z_{K_0^\ast}\,w_{K^\ast}\,h_1^C\,\phi_C\,\chi_{C0}\overline{K}^{\ast0\mu}_{0}\partial_\mu
K^{\ast0}_0\,,
\end{align}
which obtain from the corresponding interaction Lagrangian
\begin{align}
\mathcal{L}_{\chi_{C0}K^{\ast}
\overline{K}^{\ast}_{0}}=Z_{K_0^\ast}\,w_{K^\ast}\,h_1^C\,\phi_C\,\chi_{C0}
(\overline{K}^{\ast0\mu}_{0}\partial_\mu K^{\ast0}_0-\partial_\mu
K^{\ast-}_0 K^{\ast+\mu} +\partial_\mu K^{\ast+}_0
K^{\ast-\mu}-\partial_\mu\overline{K}^{\ast0\mu}_{0}
K^{\ast0\mu})\,.
\end{align}
We compute the decay width as
\begin{equation}
\Gamma_{\chi_{C0}\rightarrow K^{\ast0}
\overline{K}^{\ast0}_{0}}=\frac{|\overrightarrow{k}_1|}{8\pi\,m_{\chi_{C0}}^2}w_{K^\ast}^2\,Z_{K^\ast}(h_1^C)^2\phi_C^2\left\vert-m_{K^{\ast0}}^2+\frac{(m_{\chi_{C0}}^2-m_{K^{\ast0}}^2-m^2_{\overline{K}^{\ast0}_{0}})^2}{4m^2_{\overline{K}^{\ast0}_{0}}}
\right\vert\,.
\end{equation}
%where
%\begin{equation}
%|\overrightarrow{k}_1|=\frac{1}{2m_{\chi_{C0}}}\bigg[m_{\chi_{C0}}^4+(m_{K^{\ast0}}^2-m_{\overline{K}^{\ast0}_{0}}^2)^2-2(m_{K^{\ast0}}^2+m_{\overline{K}^{\ast0}_{0}}^2)m_{\chi_{C0}}^2\bigg]^{1/2}\,.
%\end{equation}

Note that we considered only the decay channel
$\chi_{C0}K^{\ast0} \overline{K}^{\ast0}_{0}$ because the other
decay channels contribute the same of isospin symmetry
reasons. Thus,
\begin{align}
\Gamma_{\chi_{C0}\rightarrow K^{\ast}
\overline{K}^{\ast}_{0}}=\Gamma_{\chi_{C0}\rightarrow K^{\ast0}
\overline{K}^{\ast0}_{0}}+\Gamma_{\chi_{C0}\rightarrow K^{\ast+}
K^{\ast-}_{0}}+\Gamma_{\chi_{C0}\rightarrow K^{\ast-}
K^{\ast+}_{0}}+\Gamma_{\chi_{C0}\rightarrow K^{\ast0}_0
\overline{K}^{\ast0}}\,.
\end{align}

\vspace*{5mm}
\textbf{Decay channels $\chi_{C0}\rightarrow \eta,\,\eta'$}\\

The corresponding interaction Lagrangian of $\chi_{C0}$ with the $\eta'$ and the $\eta$ resonances reads
\begin{align}
\mathcal{L}_{\chi_{C0}\eta_N\eta_N,\eta_S\eta_S,\eta_N\eta_S}=&
(-\lambda_1^C-\frac{1}{2}c\phi_N^2\phi_S^2)Z^2_{\eta_N}\,\phi_C\,\chi_{C0}\eta_N^2+\frac{1}{2}h_1^C\,w_{f_{1N}}^2\,Z_{\eta_N}^2\chi_{C0}\,\partial_\mu\eta_N\partial^\mu\eta_N\nonumber\\&
+
(-\lambda_1^C-\frac{1}{8}c\phi_N^4)Z^2_{\eta_S}\,\phi_C\,\chi_{C0}\eta_S^2+\frac{1}{2}h_1^C\,w_{f_{1S}}^2Z_{\eta_S}^2\chi_{C0}\partial_\mu\eta_S\partial^\mu\eta_S\nonumber\\&
-\frac{1}{2}\phi_N^3\phi_C\phi_S\,Z_{\eta_N}\,Z_{\eta_S}\,\eta_N\eta_S\,.\label{chietas}
\end{align}
By using Eq. (\ref{mixetas}), the interaction Lagrangian
(\ref{chietas}) will transform to a Lagrangian which describes the interaction of $\chi_{C0}$
with $\eta$ and $\eta'$,
\begin{align}
\mathcal{L}_{\chi_{C0}\eta^2,\eta'^2,\eta\eta'}&=
[-\lambda_1^C(Z^2_{\eta_N}\cos^2\varphi_\eta+Z^2_{\eta_S}\sin^2\varphi_\eta)-\frac{1}{2}c\phi_N^2(\phi_S^2\,Z^2_{\eta_N}\,\cos^2\varphi_\eta\nonumber\\
&+\frac{1}{4}\phi_N^2Z^2_{\eta_S}\sin^2\varphi_\eta+\phi_N\phi_S\,Z_{\eta_N}Z_{\eta_S}
\sin\varphi_\eta \cos\varphi_\eta)]\phi_C\chi_{C0}\eta^2\nonumber\\
&+\bigg[\frac{1}{2}h_1^C\phi_C(w^2_{f_{1N}}Z_{\eta_N}^2\cos^2\varphi_\eta+w^2_{f_{1S}}Z_{\eta_S}^2\sin^2\varphi_\eta)\bigg]\chi_{C0}\partial_\mu
\eta\partial^\mu\eta\nonumber\\
&[-\lambda_1^C(Z^2_{\eta_N}\sin^2\varphi_\eta+Z^2_{\eta_S}\cos^2\varphi_\eta)-\frac{1}{2}c\phi_N^2(\phi_S^2\,Z^2_{\eta_N}\,\sin^2\varphi_\eta\nonumber\\
&+\frac{1}{4}\phi_N^2Z^2_{\eta_S}\cos^2\varphi_\eta-\phi_N\phi_S\,Z_{\eta_N}Z_{\eta_S}
\sin\varphi_\eta \cos\varphi_\eta)]\phi_C\chi_{C0}\eta'^2\nonumber\\
&+\bigg[\frac{1}{2}h_1^C\phi_C(w^2_{f_{1N}}Z_{\eta_N}^2\sin^2\varphi_\eta+w^2_{f_{1S}}Z_{\eta_S}^2\cos^2\varphi_\eta)\bigg]\chi_{C0}\partial_\mu
\eta'\partial^\mu\eta'\nonumber\\
&+[-2\lambda_1^C(-Z^2_{\eta_N}+Z^2_{\eta_S})\sin\varphi_\eta
\cos\varphi_\eta)\nonumber\\
&+\frac{1}{2}c\phi_N^2(2\phi_S^2\,Z^2_{\eta_N}-\frac{1}{2}\phi_N^2\,Z_{\eta_S}^2)\cos\varphi_\eta
\sin\varphi_\eta)\nonumber\\&-
 \frac{1}{2}c\phi_N^3\phi_S\,Z_{\eta_N}Z_{\eta_S}(\cos^2\varphi_\eta-
\sin^2\varphi_\eta \cos\varphi_\eta)] \phi_C\chi_{C0}\eta\eta'\nonumber\\
&+h_1^C\phi_C\,\cos\varphi_\eta
\sin\varphi_\eta(w^2_{f_{1S}}Z_{\eta_S}^2-w^2_{f_{1N}}Z_{\eta_N}^2)\chi_{C0}\partial_\mu
\eta\partial^\mu\eta'\,,
\end{align}
which contains three different decay channels,
$\chi_{C0}\rightarrow\eta\eta$, $\chi_{C0}\rightarrow\eta'\eta'$,
and $\chi_{C0}\rightarrow\eta\eta'$, with the following vertices
\begin{align}
A_{\chi_{C0}\eta\eta}=&-\lambda_1^C\phi_C(Z^2_{\eta_N}\cos^2\varphi_\eta+Z^2_{\eta_S}\sin^2\varphi_\eta)\nonumber\\
&-\frac{1}{2}c\phi_N^2\phi_C(\phi_S^2\,Z^2_{\eta_N}\,\cos^2\varphi_\eta+\frac{1}{4}\phi_N^2Z^2_{\eta_S}\sin^2\varphi_\eta+\phi_N\phi_S\,Z_{\eta_N}Z_{\eta_S}
\sin\varphi_\eta \cos\varphi_\eta)\,,\\
B_{\chi_{C0}\eta\eta}&=\frac{1}{2}h_1^C\phi_C(w^2_{f_{1N}}Z_{\eta_N}^2\cos^2\varphi_\eta+w^2_{f_{1S}}Z_{\eta_S}^2\sin^2\varphi_\eta)\,,
\end{align}
\begin{align}
A_{\chi_{C0}\eta'\eta'}=&-\lambda_1^C\phi_C(Z^2_{\eta_N}\sin^2\varphi_\eta+Z^2_{\eta_S}\cos^2\varphi_\eta)\nonumber\\
&-\frac{1}{2}c\phi_N^2\phi_C(\phi_S^2\,Z^2_{\eta_N}\,\sin^2\varphi_\eta+\frac{1}{4}\phi_N^2Z^2_{\eta_S}\cos^2\varphi_\eta-\phi_N\phi_S\,Z_{\eta_N}Z_{\eta_S}
\sin\varphi_\eta \cos\varphi_\eta)\,,\\
B_{\chi_{C0}\eta'\eta'}&=\frac{1}{2}h_1^C\phi_C(w^2_{f_{1N}}Z_{\eta_N}^2\sin^2\varphi_\eta+w^2_{f_{1S}}Z_{\eta_S}^2\cos^2\varphi_\eta)\,,
\end{align}
\begin{align}
A_{\chi_{C0}\eta\eta'}=&-2\lambda_1^C\phi_C(-Z^2_{\eta_N}+Z^2_{\eta_S})\sin\varphi_\eta
\cos\varphi_\eta+\frac{1}{2}c\phi_N^2\phi_C(2\phi_S^2\,Z^2_{\eta_N}-\frac{1}{2}\phi_N^2\,Z_{\eta_S}^2)\cos\varphi_\eta
\sin\varphi_\eta\nonumber\\&-
 \frac{1}{2}c\phi_N^3\phi_S\phi_C\,Z_{\eta_N}Z_{\eta_S}(\cos^2\varphi_\eta-
\sin^2\varphi_\eta \cos\varphi_\eta)\,,\\
B_{\chi_{C0}\eta\eta'}&=h_1^C\phi_C\,\cos\varphi_\eta
\sin\varphi_\eta(w^2_{f_{1S}}Z_{\eta_S}^2-w^2_{f_{1N}}Z_{\eta_N}^2)\,.
\end{align}
Let us firstly consider the channel
$\chi_{C0}\rightarrow\eta\eta$. We denote the momenta of the two
outgoing $\eta$ particles as $P_1$ and $P_2$, and $P$ denotes
the momentum of the decaying $\chi_{C0}$ particle. Given that our
particles are on shell, we obtain
\begin{equation}
P_1\cdot P_2=\frac{P^2-P_1^2-P_2^2}{2}=\frac{m_{\chi_{C0}}^2-2m_{\eta}^2}{2}\,.
\end{equation}
After replacing $\partial_\mu\rightarrow+iP^\mu$ for the
outgoing particles, one obtains the decay amplitude as
\begin{equation}
-iM_{\chi_{C0}\rightarrow\eta\eta}=i\bigg[A_{\chi_{C0}\eta\eta}-B_{\chi_{C0}\eta\eta}\frac{m_{\chi_{C0}}^2-2m_{\eta}^2}{2}\bigg]\,.
\end{equation}
Then the decay width is
\begin{equation}
\Gamma_{\chi_{C0}\rightarrow
\eta\eta}=2\frac{|\overrightarrow{k}_1|}{8\pi\,m_{\chi_{C0}}^2}\left\vert\overline{-i\mathcal{M}_{\chi_{C0}\rightarrow\eta\eta}}\right\vert^{2}\,.
\end{equation}
%where
%\begin{equation}
%|\overrightarrow{k}_1|=\frac{1}{2m_{\chi_{C0}}}\bigg[m_{\chi_{C0}}^4-4m_{\eta}^2m_{\chi_{C0}}^2\bigg]^{1/2}\,.
%\end{equation}
Similarly, the decay width of $\chi_{C0}$ into $\eta'\eta'$ is obtained
as
\begin{equation}
\Gamma_{\chi_{C0}\rightarrow
\eta'\eta'}=2\frac{|\overrightarrow{k}_1|}{8\pi\,m_{\chi_{C0}}^2}\left\vert
A_{\chi_{C0}\eta'\eta'}-B_{\chi_{C0}\eta'\eta'}\frac{m_{\chi_{C0}}^2-2m_{\eta'}^2}{2}\right\vert^{2}\,.
\end{equation}
%where
%\begin{equation}
%|\overrightarrow{k}_1|=\frac{1}{2m_{\chi_{C0}}}\bigg[m_{\chi_{C0}}^4-4m_{\eta'}^2m_{\chi_{C0}}^2\bigg]^{1/2}\,.
%\end{equation}
In a similar way, the decay width of $\chi_{C0}$ into
$\eta\eta'$ can be obtained as
\begin{equation}
\Gamma_{\chi_{C0}\rightarrow
\eta\eta'}=\frac{|\overrightarrow{k}_1|}{8\pi\,m_{\chi_{C0}}^2}\left\vert
A_{\chi_{C0}\eta\eta'}-B_{\chi_{C0}\eta\eta'}\frac{m_{\chi_{C0}}^2-m_\eta^2-m_{\eta'}^2}{2}\right\vert^{2}\,.
\end{equation}
%where
%\begin{equation}
%|\overrightarrow{k}_1|=\frac{1}{2m_{\chi_{C0}}}\bigg[m_{\chi_{C0}}^4+(m_\eta^2+m_{\eta'}^2)^2-2(m_\eta^2+m_{\eta'}^2)m_{\chi_{C0}}^2\bigg]^{1/2}\,.
%\end{equation}

\vspace*{5mm}
\textbf{Decay channels $\chi_{C0}\rightarrow f_0f_0$}\\

The corresponding interaction Lagrangian is extracted from the Lagrangian (\ref{Lagc}) 
\begin{align}\label{f0f0}
\mathcal{L}_{\chi_{C0}
f_0 f_0}=-\lambda_1^C\,\phi_C\,\chi_{C0}(\sigma_N^2+\sigma_S^2)-\frac{m_0^2}{G_0^2}\,\phi_C \,\chi_{C0}\,G^2\,.
\end{align}
By using the mixing matrices (\ref{scalmix1}), (\ref{U2}) and (\ref{U3}), we obtain the interaction Lagrangian (\ref{f0f0}) as a function of all the following channels:
   $$\chi_{C0}\rightarrow f_0(1370)f_0(1370),\,\chi_{C0}\rightarrow f_0(1500)f_0(1500),$$ 
   $$\chi_{C0}\rightarrow f_0(1370)f_0(1500),\,\,\chi_{C0}\rightarrow f_0(1370)f_0(1710),$$
    $$\chi_{C0}\rightarrow f_0(1500)f_0(1710),\,\,$$

Then, we compute the decay widths for all these channels by using the formula of the two-body decay (\ref{B1}).

\subsection{Three-body decay rates for $\chi_{C0}$}

\textbf{Decay channel $\chi_{C0}\rightarrow K^\ast_0K\eta,\eta'$}

The corresponding interaction Lagrangian can be obtained from the Lagrangian (\ref{Lagc}) as
\begin{align}
\mathcal{L}_{\chi_{C0}K^\ast_0K\eta_N,\eta_S}=&\frac{1}{\sqrt{2}}(c-\frac{1}{2}\phi_C^2 \delta \widetilde{c})Z_K\,Z_{K^\ast_0}\,Z_{\eta_N}\phi_N^2\phi_S\phi_C\chi_{C0}\eta_N(K^{\ast0}_0\overline{K}^0
+\overline{K}^{\ast0}_0 K^0+K^{\ast-}_0K^+
+K^{\ast+}_0K^-)\nonumber\\&+\frac{\sqrt{2}}{4}(c-\frac{1}{2}\phi_C^2 \delta \widetilde{c})Z_K\,Z_{K^\ast_0}\,Z_{\eta_S}\phi_N^3\phi_C\chi_{C0}\eta_S(K^{\ast0}_0\overline{K}^0
+\overline{K}^{\ast0}_0 K^0+K^{\ast-}_0K^+ +K^{\ast+}_0K^-)\,.\label{chiKetKs}
\end{align}
Using Eq. (\ref{mixetas}), the interaction Lagrangian (\ref{chiKetKs}) can be written as
\begin{align}
\mathcal{L}_{\chi_{C0}K^\ast_0K\eta,\eta'}=\frac{1}{\sqrt{2}}(c-\frac{1}{2}\phi_C^2 \delta \widetilde{c})\phi_C\phi_N^2 Z_K Z_{K^\ast_0}\chi_{co}\bigg[&(\phi_S Z_{\eta_N} \cos\varphi_\eta+\frac{1}{2}\phi_N Z_{\eta_S} \sin\varphi_\eta)\eta\nonumber\\&
-(\phi_S Z_{\eta_N} \sin\varphi_\eta-\frac{1}{2}\phi_N Z_{\eta_S} \cos\varphi_\eta)\eta'\bigg]\nonumber\\&
\times\big[K^{\ast0}_0\overline{K}^0
+\overline{K}^{\ast0}_0 K^0+K^{\ast-}_0K^+ +K^{\ast+}_0K^-\big]\,.\label{chiKetKs1}
\end{align}
Consequently, the amplitude decay for the decay channels $\chi_{C0} \rightarrow K^\ast_0K\eta$ and $\chi_{C0}\rightarrow K^\ast_0K\eta'$ can be obtain as
\begin{equation}
-i\mathcal{M}_{\chi_{C0}\rightarrow K^\ast_0K\eta}=-i\frac{1}{\sqrt{2}}(c-\frac{1}{2}\phi_C^2 \delta \widetilde{c})\phi_C\phi_N^2 Z_K Z_{K^\ast_0}(\phi_S Z_{\eta_N} \cos\varphi_\eta+\frac{1}{2}\phi_N Z_{\eta_S} \sin\varphi_\eta)\,,
\end{equation}
and 
\begin{equation}
-i\mathcal{M}_{\chi_{C0}\rightarrow K^\ast_0K\eta'}=\frac{1}{\sqrt{2}}(c-\frac{1}{2}\phi_C^2 \delta \widetilde{c})\phi_C\phi_N^2 Z_K Z_{K^\ast_0}(-\phi_S Z_{\eta_N} \sin\varphi_\eta+\frac{1}{2}\phi_N Z_{\eta_S} \cos\varphi_\eta)\,,
\end{equation}
which are used to compute $\Gamma_{\chi_{C0}\rightarrow K^\ast_0K\eta}$ and  $\Gamma_{\chi_{C0}\rightarrow K^\ast_0K\eta'}$ by Eq.(\ref{3bodydecay}).\\

\vspace*{5mm}
\textbf{Decay channel $\chi_{C0}\rightarrow f_0\eta,\eta'$}\\

The corresponding interaction Lagrangian is extracted from the Lagrangian (\ref{Lagc}) and given by
\begin{align}
\mathcal{L}_{\chi_{C0}\sigma_{N,S}\eta_{NS}}=&-\frac{3}{2}(c-\frac{1}{2}\phi_C^2 \delta \widetilde{c})\,Z_{\eta_N}Z_{\eta_S}\phi_N^2\phi_S\phi_C\chi_{C0}\sigma_N\eta_N\eta_S-c\,Z^2_{\eta_N}\phi_N^2\phi_S\phi_C\chi_{C0}\sigma_S\eta_N^2\nonumber\\&-(c-\frac{1}{2}\phi_C^2 \delta \widetilde{c})\,Z^2_{\eta_N}\phi_N^2\phi^2_S\phi_C\chi_{C0}\sigma_N\eta^2_N-\frac{1}{2}(c-\frac{1}{2}\phi_C^2 \delta \widetilde{c})\,Z^2_{\eta_S}\phi_N^3\phi_C\chi_{C0}\sigma_N\eta^2_S\nonumber\\&-\frac{1}{2}(c-\frac{1}{2}\phi_C^2 \delta \widetilde{c})Z_{\eta_N},Z_{\eta_S}\phi_N^3\phi_C\chi_{C0}\sigma_S\eta_S\eta_N\,.\label{chif0eta}
\end{align}
We get the decay amplitudes for the following channels:
$$\chi_{C0}\rightarrow  f_0(1370)\eta\eta,\,\,\chi_{C0}\rightarrow  f_0(1370)\eta'\eta'$$
$$\chi_{C0}\rightarrow  f_0(1370)\eta\eta',\,\,\chi_{C0}\rightarrow  f_0(1500)\eta\eta$$
$$\chi_{C0}\rightarrow  f_0(1500)\eta\eta',\,\,\chi_{C0}\rightarrow  f_0(1710)\eta\eta$$
$$\chi_{C0}\rightarrow  f_0(1710)\eta\eta',\,\,$$
which used in Eq.(\ref{3bodydecay}) to compute the decay widths for all these channels.

\section{Decay rates for $\eta_C$ \label{Sec3}}

We present the explicit expressions for the two- and three-body decay
rates for the pseudoscalar hidden-charmed meson $\eta_C$.

The chiral Lagrangian contains the tree-level vertices for the
decay processes of the pseudoscalar $\eta_C$ into (pseudo)scalar
mesons, through the chiral anomaly term
\begin{equation}
\mathcal{L}_{U(1)_A}=c(det\Phi-det\Phi^{\dagger})^{2}-\delta \widetilde{c}(det\Phi-det\Phi^{\dagger})^{2}\mathrm{Tr}(\mathbb{P}_C\Phi^{\dagger}\mathbb{P}_C\Phi)
\text{ .}
\label{intlag3}%
\end{equation}
The terms in the Lagrangian (\ref{intlag3}) which correspond to decay processes of $\eta_C$ read
\begin{align}
\mathcal{L}_{\eta_{C}}=&\frac{1}{8}(c-\frac{1}{2}\phi_C^2 \delta \widetilde{c})\phi_N^2\phi_C\,Z_{\eta_C}\eta_C\{\sqrt{2}\phi_S\phi_N Z_K
Z_{K^\ast_0}(K^{\ast0}_0\overline{K}^0+\overline{K}^{\ast0}_0 K^0+K^{\ast-}_0 K^{+}+K^{\ast+}_0 K^-)\nonumber\\
&+2 Z_\pi \phi_S^2 (a_0^{0} \pi^{0 }+ a_0^{+} \pi^{-} + a_0^{-}\pi^{+})-4 \phi_N \phi_S (Z_{\eta_S} \eta_S \sigma_N + Z_{\eta_N} \eta_N \sigma_S)\nonumber\\
&-6\phi_S^2\,Z_{\eta_N}\eta_N\,\sigma_N-\phi_N^2\eta_S\sigma_S\,Z_{\eta_S}+2\phi_N\,Z^2_{\eta_S}\,Z_{\eta_N}\eta_S^2\eta_N +6\phi_S\,Z_{\eta_S}\,Z_{\eta_N}^2\eta_N^2\eta_S \nonumber\\
&-\sqrt{2}\phi_N\,Z_{\eta_S}\,Z_K^2(\overline{K}^0K^0+K^-K^+)\,\eta_S
-3\sqrt{2}\phi_S\,Z_{\eta_N}Z_K^2(\overline{K}^0K^0+K^-K^+)\eta_N\nonumber\\
&+\sqrt{2}\phi_S\,Z_\pi\,Z_K^2\bigg[\sqrt{2}(\overline{K}^0 K^+ \pi^- + K^0 K^- \pi^+)-(K^0 \overline{K}^0 -K^{-}K^{+})\pi^0\bigg]\nonumber\\
&-2\phi_S\eta_S\,Z_{\eta_S}Z_\pi^2({\pi^0}^2+2\pi^-\pi^+)\}\,.\label{intLagofetadecay}
\end{align}

\subsection{Two-body decay expressions for $\eta_C$}

The explicit expressions for the two-body decay
widths of $\eta_C$ are given by \\

\textbf{Decay channel $\eta_C\rightarrow\overline{K}^\ast_0K$}\\

The corresponding interaction Lagrangian can be obtained from the Lagrangian ( \ref{intLagofetadecay}) as
\begin{align}
\mathcal{L}_{\eta_{C}}=&\frac{\sqrt{2}}{8}(c-\frac{1}{2}\phi_C^2 \delta \widetilde{c})\phi_N^3\phi_S \phi_C\,Z_K
Z_{K^\ast_0}\,Z_{\eta_C}\eta_C(K^{\ast0}_0\overline{K}^0+\overline{K}^{\ast0}_0 K^0+K^{\ast-}_0 K^{+}+K^{\ast+}_0 K^-)\,.
\end{align}
The decay width is obtained as 
\begin{equation}
\Gamma_{\eta_C\rightarrow
\overline{K}^\ast_0K}=\frac{|\overrightarrow{k}_1|}{68\pi\,m_{\eta_C}^2}(c-\frac{1}{2}\phi_C^2 \delta \widetilde{c})^2\,\phi_N^6\,\phi_S^2\,\phi_C^2\,Z_K^2\,Z_{K^\ast_0}^2\,Z_{\eta_C}^2\,.
\end{equation}
%with
%\begin{equation}
%|\overrightarrow{k}_1|=\frac{1}{2m_{\eta_C}}\bigg[m_{\eta_C}^4+(m_K^2-m_{K_0^\ast}^2)^2-2(m_{K^\ast_0}^2+m_K^2)m_{\eta_C}^2\bigg]^{1/2}\,.
%\end{equation}

\textbf{Decay channel $\eta_C\rightarrow a_0\,\pi$}\\

The corresponding interaction Lagrangian is extracted as 
\begin{align}
\mathcal{L}_{\eta_{C}}=&\frac{1}{4}(c-\frac{1}{2}\phi_C^2 \delta \widetilde{c})\phi_N^2\phi_S^2\phi_C\,Z_\pi\,Z_{\eta_C}\eta_C\,(a_0^{0} \pi^{0 }+ a_0^{+} \pi^{-} + a_0^{-}\pi^{+})\,.
\end{align}
The decay width is obtained as

\begin{equation}
\Gamma_{\eta_C\rightarrow
a_0\,\pi}=3\frac{|\overrightarrow{k}_1|}{128\pi\,m_{\eta_C}^2}(c-\frac{1}{2}\phi_C^2 \delta \widetilde{c})^2\,\phi_N^4\,\phi_S^4\,\phi_C^2\,Z_\pi^2\,Z_{\eta_C}^2\,.
\end{equation}
%with
%\begin{equation}
%|\overrightarrow{k}_1|=\frac{1}{2m_{\eta_C}}\bigg[m_{\eta_C}^4+(m_\pi^2-m_{a_0}^2)^2-2(m_{\pi}^2+m_{a_0}^2)m_{\eta_C}^2\bigg]^{1/2}\,.
%\end{equation}

\textbf{Decay channel $\eta_C\rightarrow f_0\eta,\,\eta'$}\\

The corresponding interaction Lagrangian can be obtained from the Lagrangian ( \ref{intLagofetadecay}) as

\begin{align}
\mathcal{L}_{\eta_{C}\eta \sigma }=&\frac{1}{8}(c-\frac{1}{2}\phi_C^2 \delta \widetilde{c})\phi_N^2\phi_C\,Z_{\eta_C}\eta_C\{-4 \phi_N \phi_S (Z_{\eta_S} \eta_S \sigma_N + Z_{\eta_N} \eta_N \sigma_S)\nonumber\\
&\,\,\,\,\,\,\,\,\,\,\,\,\,\,\,\,\,\,\,\,\,\,\,\,\,\,\,\,\,\,\,\,\,\,\,-6\phi_S^2\,Z_{\eta_N}\eta_N\,\sigma_N-\phi_N^2\eta_S\sigma_S\,Z_{\eta_S}\}\,.\label{etacetasigmaint} 
\end{align}
By substituting from Eqs.(\ref{mixetas}) and mixing matrices (\ref{scalmix1}),  (\ref{U2}) and (\ref{U3}), then using the two-body decay (\ref{B1}). We get the results of several scenarios of the decay widths:
 $$\Gamma_{\eta_C\rightarrow f_0(1370)\eta},\,\,\Gamma_{\eta_C\rightarrow f_0(1500)\eta},\,\,\Gamma_{\eta_C\rightarrow f_0(1710)\eta}\,,$$
$$\Gamma_{\eta_C\rightarrow f_0(1370)\eta'},\,\,\Gamma_{\eta_C\rightarrow f_0(1500)\eta'},\,\,\Gamma_{\eta_C\rightarrow f_0(1710)\eta'}\,.$$

\subsection{Three-body decay expressions for $\eta_C$}

The corresponding interaction Lagrangian, contains the three-body decay rates for the $\eta_C$ meson, is extracted as
\begin{align}
\mathcal{L}_{\eta_{C}}=&\frac{1}{8}(c-\frac{1}{2}\phi_C^2 \delta \widetilde{c})\phi_N^2\phi_C\,Z_{\eta_C}\eta_C\{2\phi_N\,Z^2_{\eta_S}\,Z_{\eta_N}\eta_S^2\eta_N +6\phi_S\,Z_{\eta_S}\,Z_{\eta_N}^2\eta_N^2\eta_S \nonumber\\
&-\sqrt{2}\phi_N\,Z_{\eta_S}\,Z_K^2(\overline{K}^0K^0+K^-K^+)\,\eta_S
-3\sqrt{2}\phi_S\,Z_{\eta_N}Z_K^2(\overline{K}^0K^0+K^-K^+)\eta_N\nonumber\\
&+\sqrt{2}\phi_S\,Z_\pi\,Z_K^2\bigg[\sqrt{2}(\overline{K}^0 K^+ \pi^- + K^0 K^- \pi^+)-(K^0 \overline{K}^0 -K^{-}K^{+})\pi^0\bigg]\nonumber\\
&-2\phi_S\eta_S\,Z_{\eta_S}Z_\pi^2({\pi^0}^2+2\pi^-\pi^+)\}\,.\label{intLageta3decay}
\end{align}
 
Using the general formula for the three-body decay
width for $\eta_C$ (\ref{3bodydecay}), which the corresponding tree-level decay amplitudes for $\eta_C$ are obtained as follows:\\

\textbf{Decay channel $\eta_C\rightarrow \eta^3$:}
$m_1=m_2=m_3=m_{\eta}$ and $S=6$
\begin{equation}
|\overline{-iM_{\eta_C\rightarrow
\eta^3}}|^2=\bigg[\frac{1}{4}(c-\frac{1}{2}\phi_C^2 \delta \widetilde{c})\phi_N^2\phi_C\, \sin\varphi_\eta
\cos\varphi_\eta(Z_{\eta_S}\phi_N
\sin\varphi_\eta+3Z_{\eta_N}\phi_S\,\cos\varphi_{\eta})Z_{\eta_C}Z_{\eta_N}Z_{\eta_S}\bigg]^2\,.
\end{equation}
$$\\$$

\textbf{Decay channel $\eta_C\rightarrow \eta'^3$:}
$m_1=m_2=m_3=m_{\eta'}$ and $S=6$
\begin{equation}
|\overline{-iM_{\eta_C\rightarrow
\eta'^3}}|^2=\bigg[\frac{1}{4}(c-\frac{1}{2}\phi_C^2 \delta \widetilde{c})\phi_N^2\phi_C\, \sin\varphi_\eta
\cos\varphi_\eta(Z_{\eta_S}\phi_N
\cos\varphi_\eta-3Z_{\eta_N}\phi_S\,\sin\varphi_{\eta})Z_{\eta_C}Z_{\eta_N}Z_{\eta_S}\bigg]^2\,.
\end{equation}
$$\\$$

\textbf{Decay channel $\eta_C\rightarrow \eta'\eta^2$:}
$m_1=m_{\eta'},\,\, m_2=m_3=m_{\eta}$ and $S=2$
\begin{align}
|\overline{-iM_{\eta_C\rightarrow
\eta'\eta^2}}|^2=\frac{1}{16}(c-\frac{1}{2}\phi_C^2 \delta \widetilde{c})^2\phi_N^4\phi_C^2\,\bigg[&\phi_N\,Z_{\eta_S}\,\sin\varphi_\eta
(2\cos^2\varphi_\eta-\sin^2\varphi_\eta)\nonumber\\&+3Z_{\eta_N}\phi_S
\cos\varphi_\eta(\cos^2\varphi_\eta-2\,\sin^2\varphi_{\eta})\bigg]^2\,.
\end{align}
$$\\$$

\textbf{Decay channel $\eta_C\rightarrow \eta'^2\eta$:}
$m_1=m_2=m_{\eta'},\,\, m_3=m_{\eta}$ and $S=2$
\begin{align}
|\overline{-iM_{\eta_C\rightarrow
\eta'^2\eta}}|^2=\frac{1}{16}(c-\frac{1}{2}\phi_C^2 \delta \widetilde{c})^2\phi_N^4\phi_C^2\,Z^2_{\eta_C}\,Z^2_{\eta_N}\,Z^2_{\eta_S}\bigg[&Z_{\eta_S}\phi_N
\cos\varphi_\eta(\cos^2\varphi_\eta-2
\sin^2\varphi_\eta)\nonumber\\&+3Z_{\eta_N}\phi_S
\sin\varphi_\eta(\cos^2\varphi_\eta-2\,\cos^2\varphi_{\eta})\bigg]^2\,.
\end{align}
$$\\$$ 
 
\textbf{Decay channel $\eta K \overline{K}$:}
$m_1=K^0,\,m_2=m_{\overline{K}^0},\,\, m_3=m_{\eta}$
\begin{align}
\Gamma_{\eta_C\rightarrow \eta K
\overline{K}}&=\Gamma_{\eta_C\rightarrow \eta K^+
K^-}+\Gamma_{\eta_C\rightarrow \eta K^0
\overline{K}^0}\nonumber\\&=2\Gamma_{\eta_C\rightarrow \eta K^0
\overline{K}^0}.
\end{align}
with the average modulus squared decay amplitude
\begin{equation}
|\overline{-iM_{\eta_C\rightarrow \eta
K\overline{K}}}|^2=\frac{1}{32}(c-\frac{1}{2}\phi_C^2 \delta \widetilde{c})^2\phi_N^4\phi_C^2\,Z^2_{\eta_C}\,Z^4_{K}\,(\phi_N\,Z_{\eta_S}\,\sin\varphi_\eta+3\phi_S\,Z_{\eta_N}\cos\varphi_\eta)^2\,.
\end{equation}
$$\\$$

\textbf{Decay channel $\eta_C\rightarrow \eta' K \overline{K}$:}
$m_1=K^0,\,m_2=m_{\overline{K}^0},\,\, m_3=m_{\eta'}$
\begin{equation}
\Gamma_{\eta_C\rightarrow \eta' K
\overline{K}}=2\Gamma_{\eta_C\rightarrow \eta' K^0
\overline{K}^0}.
\end{equation}
The average modulus squared decay amplitude for this process reads
\begin{equation}
|\overline{-iM_{\eta_C\rightarrow \eta'
K\overline{K}}}|^2=\frac{1}{32}(c-\frac{1}{2}\phi_C^2 \delta \widetilde{c})^2\phi_N^4\phi^2_C\,Z^2_{\eta_C}\,Z^4_{K}\,(\phi_N\,Z_{\eta_S}\,\cos\varphi_\eta-3\phi_S\,Z_{\eta_N}\sin\varphi_\eta)^2\,.
\end{equation}
$$\\$$

\textbf{Decay channel $\eta_C\rightarrow \eta \pi\pi$:}
$m_1=\eta,\,m_2=m_3=m_{\pi^0}$ and $S=2$
\begin{equation}
\Gamma_{\eta_C\rightarrow \eta
\pi\pi}=3\Gamma_{\eta_C\rightarrow \eta \pi^0 \pi^0}\,,
\end{equation}
where the average modulus squared decay amplitude for this process
 is obtained from the Lagrangian (\ref{intLageta3decay}) as
\begin{equation}
|\overline{-iM_{\eta_C\rightarrow \eta \pi\pi
}}|^2=\frac{1}{16}(c-\frac{1}{2}\phi_C^2 \delta \widetilde{c})^2\phi_N^4\phi_S^2\,\phi_C^2\,Z^2_{\eta_C}\,Z^2_{\eta_S}\,Z^4_{\pi}\,\sin^2\varphi_\eta\,.
\end{equation}
$$\\$$

\textbf{Decay channel $\eta_C\rightarrow \eta' \pi\pi$:}
$m_1=\eta',\,m_2=m_3=m_{\pi^0}$ and $S=2$
\begin{equation}
\Gamma_{\eta_C\rightarrow \eta'
\pi\pi}=3\Gamma_{\eta_C\rightarrow \eta' \pi^0 \pi^0}\,,
\end{equation}
where the average modulus squared decay amplitude for this process is
\begin{equation}
|\overline{-iM_{\eta_C\rightarrow \eta' \pi\pi
}}|^2=\frac{1}{16}(c-\frac{1}{2}\phi_C^2 \delta \widetilde{c})^2\phi_N^4\phi_S^2\,\phi_C^2\,Z_{\eta_C}^2\,Z_{\eta_S}^2\,Z^4_{\pi}\,\cos^2\varphi_\eta\,.
\end{equation}
$$\\$$

\textbf{Decay channel $\eta_C\rightarrow KK\pi$:}
$m_1=K^+,\,m_2=K^-,\,m_3=m_{\pi^0}$ and $S=2$
\begin{align}
\Gamma_{\eta_C\rightarrow KK\pi}&=\Gamma_{\eta_C\rightarrow
K^+K^-\pi^0}+\Gamma_{\eta_C\rightarrow
K^0\overline{K}^0\pi^0}+\Gamma_{\eta_C\rightarrow
\overline{K}^0K^+\pi^-}+\Gamma_{\eta_C\rightarrow K^0K^-\pi^+}\nonumber\\&
=4\Gamma_{\eta_C\rightarrow K^+K^-\pi^0}.
\end{align}

with the average modulus squared decay amplitude
\begin{equation}
|\overline{-iM_{\eta_C\rightarrow K^+K^-\pi^0
}}|^2=\frac{1}{32}(c-\frac{1}{2}\phi_C^2 \delta \widetilde{c})^2\phi_N^4\phi_S^2\,\phi_C^2\,Z^2_{\eta_C}\,Z_{\eta_K}^4\,Z_{\pi}^2\,.
\end{equation}

\bibliography{charm}

\end{document}